\documentclass[a4paper,fleqn]{cas-dc}

\usepackage[numbers]{natbib}
\usepackage{float}	
\usepackage{subfig}
\usepackage{pifont}
\usepackage{array}
\usepackage{multirow}
\usepackage{longtable}
\usepackage{algorithm}
\usepackage{algorithmic}
\usepackage{epstopdf}

\newcommand{\QQ}{\mathcal Q}
\newcommand{\UU}{\mathcal U}

\def\tsc#1{\csdef{#1}{\textsc{\lowercase{#1}}\xspace}}
\tsc{WGM}
\tsc{QE}
\tsc{EP}
\tsc{PMS}
\tsc{BEC}
\tsc{DE}
\begin{document}
\let\WriteBookmarks\relax
\def\floatpagepagefraction{1}
\def\textpagefraction{.001}
\shorttitle{Survey on DNN Watermarking}
\shortauthors{Yue Li et~al.}

\title [mode = title]{A survey of deep neural network watermarking techniques}                      
%\tnotemark[1,2]

%\tnotetext[1]{This document is the results of the research project funded by the National Science Foundation.}

%\tnotetext[2]{The second title footnote which is a longer text matter to fill through the whole text width and overflow into another line in the footnotes area of the first page.}

\author[1]{Yue Li}%[type=editor,
                       % auid=000,bioid=1,
                       % prefix=Sir,
                        %role=Researcher,
                       % orcid=0000-0001-7511-2910]
\ead{liyue859000040@my.swjtu.edu.cn}
%\credit{Conceptualization of this study, Methodology, Software}
\address[1]{School of Information Science \& Technology, Southwest Jiaotong University, 611756 Chengdu, China}

\author[2]{Hongxia Wang} %[style=chinese]
\cormark[1]
\ead{hxwang@scu.edu.cn}
\address[2]{School of Cyber Science and Engineering, Sichuan University, 610065 Chengdu, China}
%\credit{Paper Writing}

\author[3]{Mauro Barni}%[
   %role=Co-ordinator,
   %suffix=Jr,
\ead{barni@dii.unisi.it}
%\credit{Data curation, Writing - Original draft preparation}
\address[3]{Dept. Information Engineering and Mathematics, University of Siena, 53100 Siena, ITALY}

\cortext[cor1]{Corresponding author}
%\cortext[cor2]{Principal corresponding author}
%\fntext[fn1]{This is the first author footnote. but is common to third author as well.}
%\fntext[fn2]{Another author footnote, this is a very long footnote and it should be a really long footnote. But this footnote is not yet sufficiently long enough to make two lines of footnote text.}

%\nonumnote{This note has no numbers. In this work we demonstrate $a_b$ the formation Y\_1 of a new type of polariton on the interface between a cuprous oxide slab and a polystyrene micro-sphere placed on the slab. }

\begin{abstract}
Protecting the Intellectual Property Rights (IPR) associated to Deep Neural Networks (DNNs) is a pressing need pushed by the high costs required to train such networks and the importance that DNNs are gaining in our society. Following its use for Multimedia (MM) IPR protection, digital watermarking has recently been considered as a mean to protect the IPR of DNNs. While DNN watermarking inherits some basic concepts and methods from MM watermarking, there are significant differences between the two application areas, calling for the adaptation of media watermarking techniques to the DNN scenario and the development of completely new methods. In this paper, we overview the most recent advances in DNN watermarking, by paying attention to cast it into the bulk of watermarking theory developed during the last two decades, while at the same time highlighting the new challenges and opportunities characterising DNN watermarking. Rather than trying to present a comprehensive description of all the methods proposed so far, we introduce a new taxonomy of DNN watermarking and present a few exemplary methods belonging to each class. We hope that this paper will inspire new research in this exciting area and will help researchers to focus on the most innovative and challenging problems in the field.
\end{abstract}

%\includegraphics{figs/grabs.pdf}

%\item Research highlights item 1
%%\item Research highlights item 3

\begin{keywords}
Intellectual property protection \sep Deep Neural Networks \sep Watermarking \sep White box vs black box watermarking \sep Watermarking and DNN backdoors\sep
\end{keywords}

\maketitle

\section{Introduction}

Deep  Neural  Networks  (DNNs)  are  increasingly  deployed and commercialised in a wide variety of real-world scenarios due  to the unprecedented performance they achieve.  Training a DNN is a very expensive process that requires: (i) the availability of massive amounts of, often proprietary, data, capturing different scenarios within the target application; ii) extensive  computational  resources; iii) the assistance of Deep Learning (DL) experts to carefully  fine-tune  the  network  topology  (e.g., the type  and  number  of  hidden layers), and correctly set the training hyper-parameters, like the learning rate, the batch size, etc.
As a consequence, high-performance DNNs should be considered as the intellectual property (IP) of the model owner and be protected accordingly. Inspired by the use of classical watermarking techniques for the protection of property rights associated to multimedia contents, DNN watermarking is receiving increasing attention, and several  works  have been published leveraging on digital watermarking  to address IP protection in the DL domain.

In the last two decades, watermarking technology has been applied to protect multimedia documents, and a vast body of literature has been developed as summarised in several books and surveys \cite{barni2004watermarking, cox2002digital, podilchuk2001digital, cox2007digital, singh2013survey, arnold2003techniques}. Watermarked contents include audio, still images, video, graphics, text and several other kinds of media \cite{luo2011surface, shehab2007watermarking, ohbuchi2002robust}. The watermark can be injected into the host document by adding to it a low-amplitude, often pseudo-random, signal, either directly in the sample domain or in a properly transformed domain. In the case of still images, for instance, the watermark can be embedded in the spatial domain by adding a low amplitude spread spectrum signal or by substituting the least significant bits (LSB) of the pixel values \cite{cox1997secure, nikolaidis1998robust, singh2012novel}. In the case of transformed domain watermarking, several reversible transforms like the Discrete Cosine Transform (DCT)~\cite{barni1998dct, lin2010improving}, the Discrete Wavelet Transform (DWT)~\cite{ganic2004robust}, or the Discrete Fourier Transform (DFT)~\cite{BBP_DWT01,tsui2008color} are widely used to embed the watermark in a robust and imperceptible way. For audio watermarking, a number of time domain methods have been proposed, including the pioneering works in \cite{bassia2001robust} and \cite{gruhl1996echo}, and several other solutions borrowed from the image watermarking field and adapted to work with audio signals~\cite{arnold2013phase,xiang2014patchwork,khaldi2012audio}. In the case of video watermarking, we can distinguish between techniques operating on the pixel values of video frames, possibly treating them as still images~\cite{seo2006image}, and techniques working in the compressed domain. In the latter case, the watermarking algorithms are tightly tied to compression standards, like, for instance, H.264/AVC \cite{li2019robust} and HEVC \cite{tew2014information}.

In all cases, the watermarking process exploits some forms of redundancy present in the host document, thanks to which the document can be modified without impairing its informative or perceptual meaning. A similar idea holds for the case of DNN watermarking. The very large number of parameters (the network weights) DNN models consist of, confers to the network a capability of analysing the input data that often exceeds the difficulty of the task the network is trained for, hence leaving many degrees of freedom in the choice of the exact weights. The weights, then, can be modified, or directly generated, in such a way to host the watermark.

\begin{figure}
	\centering
	\includegraphics[scale=.5]{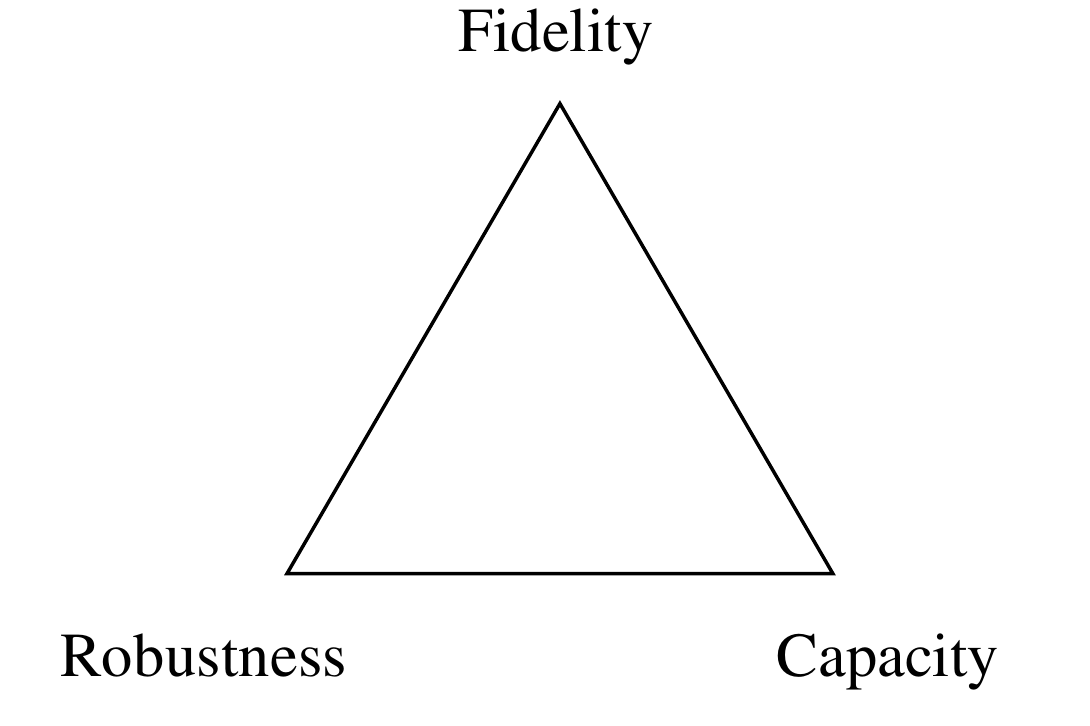}
	\caption{The watermarking trade-off triangle}
	\label{trade_off_triangle}
	{\tiny }\end{figure}

Despite the differences between watermarking techniques thought to work with different media and in different application scenarios, the requirements that any watermarking system must satisfy can be summarised by the so-called watermarking trade-off triangle shown in Figure \ref{trade_off_triangle}.
According to such a perspective, the capacity requirement, identifiable as the number of bits, or payload, conveyed by the watermark, conflicts with other two requirements, namely: i) fidelity, corresponding to watermark imperceptibility in the case of conventional media, and ii) robustness, that is the ability to recover the watermark even when the hosting document undergoes some modifications.  DNN watermarking obeys to the same principles. Here fidelity refers to the capability of the watermarked network to accomplish the task it is thought for. Robustness is related to the possibility of correctly extracting the watermark from a slightly modified version of the network model (e.g. after fine tuning), while capacity indicates the size of the payload conveyed by the watermark.

A noticeable difference between DNN watermarking and multimedia watermarking is that in the case of DNNs, the injection of the watermark can not be carried out by directly modifying the weights of the model, since in this way it would be difficult to estimate the impact of the watermark on the performance of the network. On the contrary, watermark embedding is carried out during the training process, by properly modifying the loss function used in such a phase.

%At early phase, the most simple and straightforward approach for video watermarking is to consider a video as a succession of still images and to reuse an existing image watermarking scheme \cite{doerr2003guide}. Considering different demands during transmission, along with the developing of video compression standards, more and more researchers tend to investigate watermarking methodologies based on video codec standards by leveraging characters of different types of frames (I, P and B frames) in video \cite{tew2013overview}.  For audio, watermarks can be injected into audio by modifying its Fourier coefficients \cite{bassia2001robust}.

%Following the structure of classic multimedia watermarking system, DNN watermarking obey the same progress that can be further split into three main task: (i) information coding; (ii) watermark embedding; (iii) watermark recovery/verification. Using any kind of multimedia as host asset, the classic watermark embedding amounts to the definition of an insertion operator. Unlike conventional methods, directly watermarking information into DNN models without training will incur huge sacrifice on the classification accuracy of the model.  Besides, according to embedding situations, DNN watermarking has three embedding modes that classic watermarking is not capable of: train-to-embed, fine-tune-to-embed and distill-to-embed. \MB{This distinction is not clear at all, at least at this point of the paper}

Another peculiarity of DNN watermarking is where the watermark can be read from. In a first case (referred to as {\em static} watermarking), the watermark can be read directly from the network weights, in a way that is similar to conventional multimedia watermarking techniques. In other cases, though, the effect of the watermark is to alter the {\em behaviour} of the watermarked model, when the network is fed with some specific inputs. In this case, hereafter referred to as {\em dynamic} watermarking, the watermark message is read by looking at the output of the model, or the values of the intermediate activation maps, in correspondence of properly crafted inputs. This marks a significant different with respect to classical multimedia watermarking, wherein only the static modality is possible.

By recognizing the similarities and dissimilarities between multimedia and DNN watermarking, the first goal of this paper, is to provide a taxonomy of DNN watermarking techniques based on a mix of conventional classification means and some brand new perspectives. In doing so, we pay great attention to list the essential requirements that a DNN watermarking scheme must satisfy and interpret them for the different classes of watermarking techniques.  As a second goal, we review some of the most popular and best performing DNN watermarking algorithms proposed so far, to give the reader  a clear understanding of the practical challenges and opportunities of DNN watermarking.

The rest of the paper is organized as follows. In Section 2, we introduce a taxonomy of DNN watermarking comparing it with the conventional taxonomy used for multimedia watermarking. The main requirements of DNN watermarking are discussed in Section 3. In Section 4 and 5, we review, respectively, the main static and dynamic watermarking algorithm proposed so far. Section 6 is dedicated to the presentation of the most popular attacks against DNN watermarking algorithms. Eventually, in Section 7, we draw some conclusions and highlight some directions for future works.

\section{A taxonomy of DNN watermarking}
\label{sec: taxononmy}

In this section, we present a taxonomy of DNN watermarking techniques. The taxonomy takes into account the similarities and dissimilarities between multimedia and DNN watermarking, some of which have already been outlined in the introduction. To start with, we review the most important classification criteria used to categorise classic watermarking algorithms, discussing their applicability to the DNN case. Later, we introduce some unique classification criteria based on the peculiar characteristics of DNN watermarking. Eventually, we discuss the relationship between the various watermarking classes.

\subsection{Classical watermarking models}

\begin{figure}
	\centering
	\includegraphics[scale=.4]{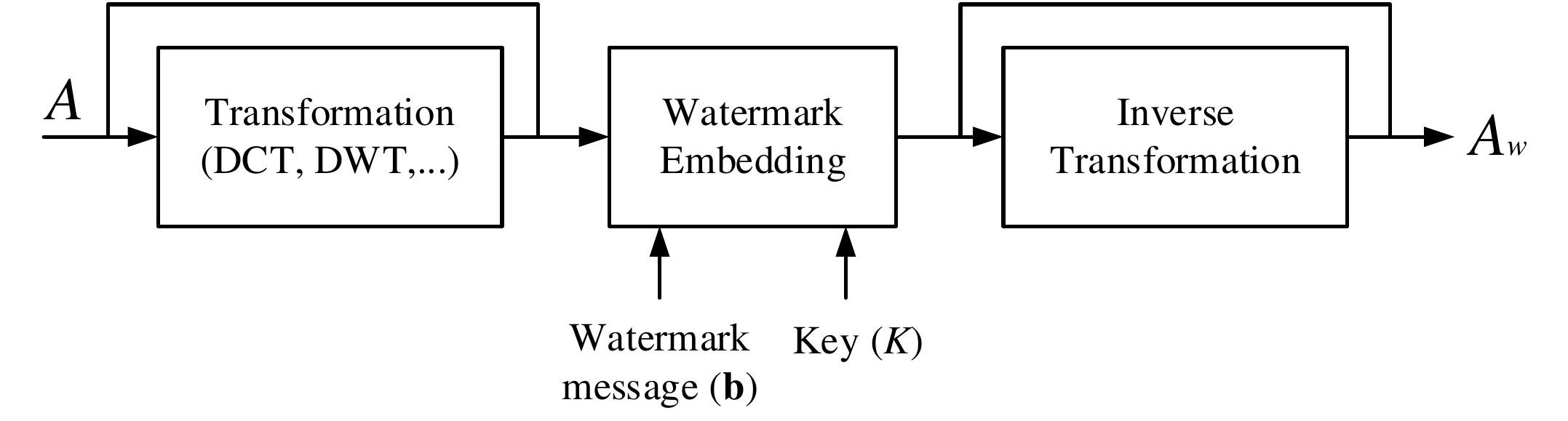}
	\caption{Overall view of classical watermark embedding.}
	\label{overall_classic_watermarking}
\end{figure}

Classically, digital watermarking aims at embedding a certain message, called watermark, within a hosting multimedia content, often, but not only, for copyright protection. Watermark injection follows the general model depicted in Figure \ref{overall_classic_watermarking}. The to-be-watermarked content $A$ is possibly transformed so that the watermark is embedded in a different domain (e.g. the DCT or wavelet domain for still images)\footnote{The direct and inverse transforms are optional steps.}, then the watermark is injected within the document, with the possible intervention of a watermarking key $K$. The watermarked content is then brought back into the original domain to obtained the watermarked document $A_w$. As simple as this scheme may seem, it contains some implicit assumptions that are not necessarily satisfied in the case of DNN watermarking: i) the existence of a to-be-watermarked content $A$ and ii) the consequent possibility of measuring the distortion between the original and the watermarked contents, $A$ and $A_w$. This is not necessarily the case with DNN watermarking, where the network weights, which are going to host the watermark, may not exist {\em per se}, since they are generated contextually to watermark embedding during the training phase. Moreover, the distortion introduced by the watermark can not be evaluated directly by measuring the difference between $A$ and $A_w$. Rather, the impact of the watermark must be measured by evaluating the performance achieved by the watermarked model.

Note that watermark embedding (and later on watermark recovery) may require the knowledge of a secret key $K$. Such a \emph{key}, whose main goal is to introduce some secrecy within the watermarking system, is commonly used to parameterize the embedding process and make the recovery of the watermark impossible for non-authorized users who do not know $K$.

% Watermarking procedure is usually divided into two steps: embedding and verification. In the embedding process, an embedding algorithm $E$ embeds pre-defined watermarks $W$ into the carrier data $C$, which is the data to be protected. After the embedding, the embedded data ($e=E(W,C)$) are stored or transmitted. During the watermark verification process, a decryption algorithm $D$ attempts to extract the watermarks $W^{\prime}$ from $e^{\prime}$. Here the input data $e^{\prime}$ may be slightly different from previously embedded data $e$ because $e$ could be modified during the transmission and distribution. Such modification could be reproduced or derived from original data $e$. Therefore, after extracting watermark $W^{\prime}$, it need to be further verified with original watermark $W$. \MB{NO. This is correct only when you pass from a multibit to a zero bit watermarking method. With zero bit watermarking, the verification may be carried out directly without that the message $W$ is first extracted. With multibit watermarking there is no need to compare the extracted bits with an original watermark} If the distance is acceptable, it is confirmed that the data is that we protected. According to this classical watermarking model, there are several classification can be further discussed.

\subsubsection{Multi-bit vs. zero-bit watermarking techniques}

\begin{figure}[htpb]
	\centering
	\subfloat[]{
		\begin{minipage}[t]{1\linewidth}
			\centering
			\includegraphics[scale=.6]{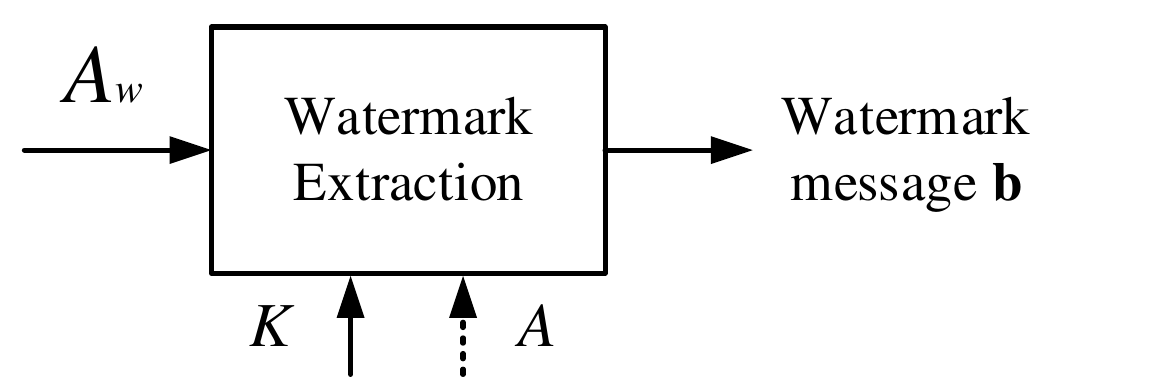}
			%			\label{static_watermarking}
		\end{minipage}
	} \\
	\subfloat[]{
		\begin{minipage}[t]{1\linewidth}
			\centering
			\includegraphics[scale=.6]{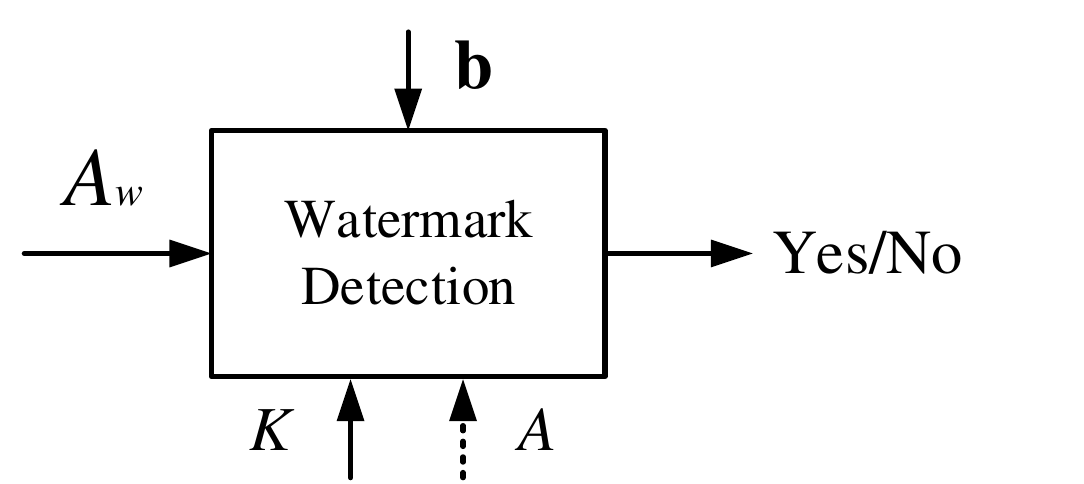}
			%			\label{dynamic_watermarking}
		\end{minipage}
	}
	\caption{Multibit (a) vs zero-bit (b) watermarking}
	\label{multi_vs_0bit}
\end{figure}

Depending on the exact content of the watermark message, and the way such a message is recovered from the host signal, we can distinguish between two main classes of algorithms: multi-bit and 0-bit watermarking. In the multi-bit case, the watermark message corresponds to a sequence of $N$ bits ${\bf b}$. Such a sequence can be read from the watermarked content, as depicted in part (a) of Figure \ref{multi_vs_0bit}. In the 0-bit case, watermark extraction corresponds to a detection task, wherein the detector is asked to decide whether a known watermark is present in the analyzed content or not. In some applications, a mixture of the two functionalities is required. The detector must first determine if a watermark is present and, if so, identify which of the $2^N$ messages is encoded. Such a detector would therefore have $2^{N}+1$ possible output values \cite{cox2007digital}. In the zero-bit case, only one possible watermark exists. In this case only $2^{0} + 1 = 2$ outputs are possible, thus justifying the \emph{zero-bit watermarking} term.

In general, multi-bit watermarking offers more flexibility\footnote{It can be shown that a multi-bit watermarking algorithm can always be transformed into a zero-bit scheme \cite{barni2004watermarking}.} and hence it can be used in a wider variety of applications, including fingerprinting \cite{kirovski2002multimedia}, error concealment \cite{adsumilli2005robust}, source tracking \cite{celik2008lookup}, labelling etc. Zero-bit watermarking generally achieves a higher robustness and is widely adopted in copyright protection  platforms, where the presence of the watermark is used as a flag to warn compliant devices that the piece of
content they are dealing with is copyrighted material \cite{furon2007constructive}. The distinction between multi-bit and zero-bit watermarking is also appropriate for DNN watermarking techniques.

\subsubsection{Robust vs. fragile watermarking}

Watermark robustness accounts for the capability of the hidden data to survive host signal manipulations, including non-malicious and malicious manipulations. Malicious manipulations precisely aim at damaging or removing the watermark, on the contrary, non-malicious manipulations indicate some unintentional or unavoidable processing that may perturb the hidden watermark, for instance, multimedia compression occurring during image or video transmission. \emph{Robust watermarking} requires the watermark to be resistant against non-malicious manipulations, whose application is implied by the normal use of the data, such as data compression, editing and cropping. A secure watermark, on the contrary, should survive also malicious manipulations.
If the hidden watermark is lost or irremediably altered as soon as any modification is applied to the host content, the watermark is said to be fragile. Data authentication is the main application for fragile watermarking, because the alteration of the watermark can be taken as an evidence that the content has been tampered with.

With regard to DNN watermarking, most of the applications considered so far call for the adoption of a robust watermark, that can survive at least some common operations like fine-tuning or nodes pruning. Watermark security may also be required in specific applications wherein the presence of an adversary aiming at watermark removal can not be ruled out.

\subsubsection{Blind vs. non-blind detection}
\label{subsub.bvnb}

From Figure \ref{multi_vs_0bit}, we can see that for both the multi-bit and the zero-bit cases, watermark recovery may require the knowledge of the non-watermarked content $A$. In classic watermarking theory, we say that a watermarking algorithm is \emph{blind}\footnote{The term oblivious may also be used.} if the watermark is recovered without resorting to the comparison between the original non-marked content and the marked one. Conversely, \emph{non-bind} watermarking algorithms use the original non-marked content to ease the extraction. As we already noted, in the case of DNN watermarking, the concept of original non-watermarked content does not apply, so the distinction between blind and non-blind watermarking does not make sense.

\subsubsection{Informed watermarking}

Informed embedding and informed coding are watermarking paradigms that have been proven to greatly improve the performance of a watermarking system \cite{miller2004applying}. Such paradigms stem from the interpretation of watermarking as a problem of channel coding with side information at the transmitter \cite{Cox99, Costa83}. In a nutshell, with informed coding each watermark message is associated to a pool of codewords (rather than to a single one), then informed watermark embedding is applied by choosing the codeword that results in the minimum distortion. Practical implementations of the informed  coding paradigm include several popular watermarking schemes like Quantization Index Modulation (QIM) \cite{Chen01}, Dither Modulation (DM) \cite{Chen98} and Scalar Costa's Scheme (SCS) \cite{Egg03}. Informed watermarking theory provides powerful means to improve the performance of watermarking algorithms, with particular reference to multi-bit watermarking, since its adoption permits to increase significantly the watermark payload without sacrificing its robustness. Nevertheless, such a theory has not been applied extensively to DNN watermarking. A DNN watermarking algorithm exploiting informed coding to increase the watermark payload is described in Section \ref{sec:static_algorithm}.
\subsection{DNN watermarking models}
\label{subsec: DNN model}

Deep learning is a machine learning framework which
automatically learns hierarchical data representation from training data without the need to resort to handcrafted feature representations \cite{goodfellow2016deep}. DNNs are the most common learning architectures on which deep learning methods are based. To be specific, a  DNN takes as input the content $x\in \mathbb{R}^{m}$ to be processed   in raw format (for instance an image or an audio signal), and maps it to the output via a parametric function, $z = F_{\theta}(x)$, where $z\in \lbrack1, n\rbrack$. The parametric function $F_{\theta} (\cdot)$ is defined by the network architecture and the
collective parameters of all the units composing it. Given the architecture, the network behaviour is determined by the values of the network parameters $\theta$ (in most cases $\theta$ corresponds to the network weights and offsets). Let $D = \{x_{i} , z_{i}\}^{T}_{i=1}$ be the training data, where $z_{i}\in \lbrack1, n\rbrack$ is the ground truth label for $x_{i}$. During the training phase, the network parameters are optimised to minimise a loss function expressing the difference between the predicted class labels and the ground truth labels. Currently, the most widely used approach for training a DNN is the back-propagation algorithm, whereby the network parameters are updated by propagating the gradient of the loss from the output layer through the entire network.
DNN watermarking is based on the observation that the huge number of parameters $\theta$ consists of, allows to slightly modify them, without that the performance of the network is degraded significantly. The parameters in $\theta$, then, can be used to encode additional information beyond what is required for the primary task of the network.

%In general, DNN watermarking model obey the same flow as classic watermarking in Figure \ref{overall_classic_watermarking}. Note that, in DNN watermarking model, the host asset $A$ substitutes DNN architectures for conventional multimedia.  In some dynamic watermarking schemes (the definition of dynamic watermarking we will discuss it in later section), the key $K$ and the watermark $\mathbf{b}$ are the same thing, in other words, a \emph{watermark key}. Since $D = \{x_{i} , z_{i}\}^{T}_{i=1}$ as an integral is used as the training data, the key and watermark are consist of $x_i$ and $z_i$ as well.

%Defining the DNN watermarking model obeys the same progress as the classic watermarking model. Inspired by \cite{barni2004watermarking}, DNN watermarking techniques can be defined by three steps:

While the classical taxonomy of watermarking algorithms described in the previous section can be largely applied also to DNN watermarking, in the following, we introduce two new characterisations that are specific for DNN watermarking.

\subsubsection{White-box vs. black-box DNN extraction}
\label{white_black}

\begin{figure}[htpb]
	\centering
	\subfloat[]{
		\begin{minipage}[t]{1\linewidth}
			\centering
			\includegraphics[scale=.32]{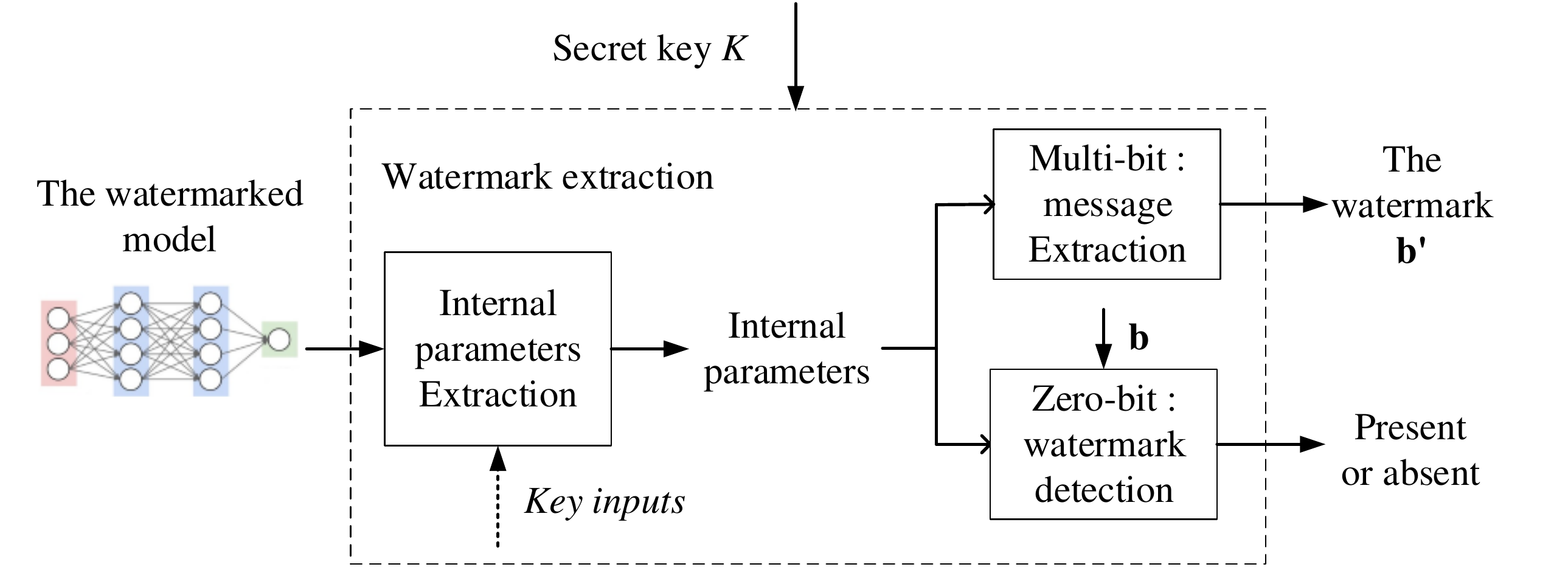}\label{white_box_watermarking}
		\end{minipage}
	} \\
	\subfloat[]{
		\begin{minipage}[t]{1\linewidth}
			\centering
			\includegraphics[scale=.33]{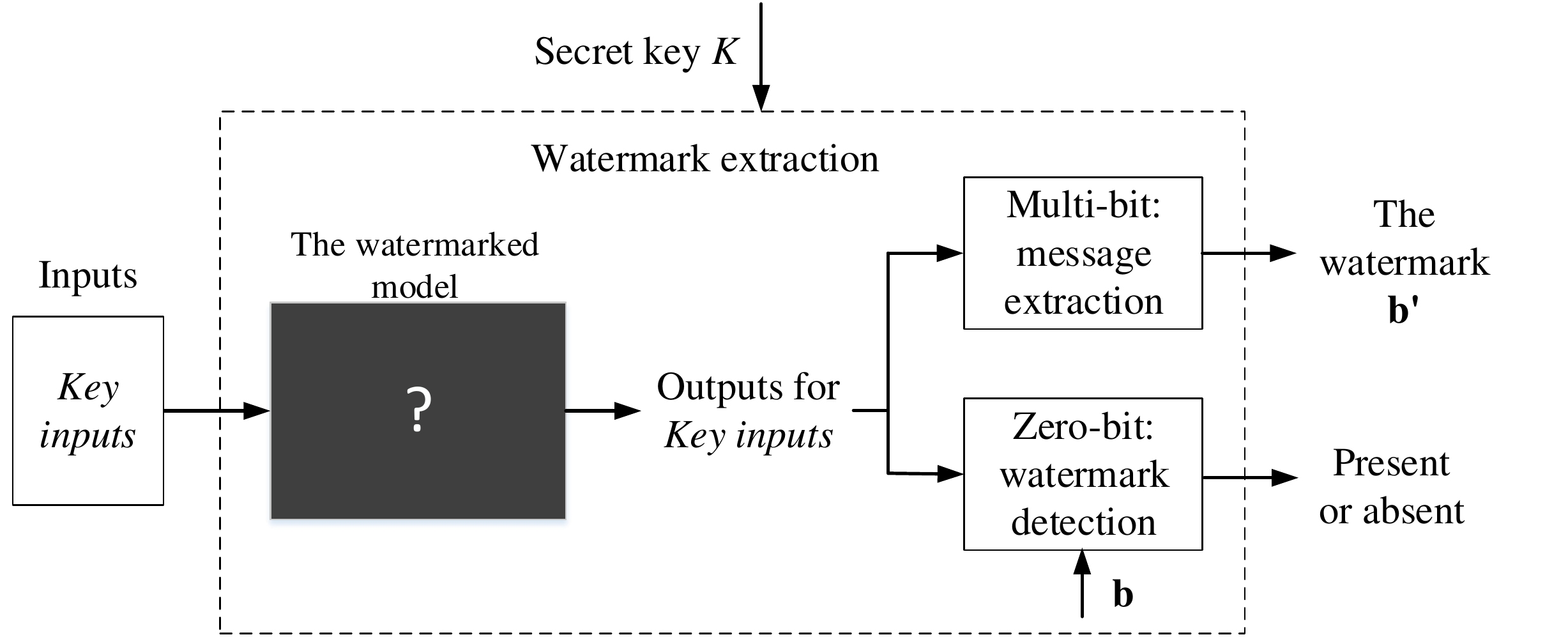}\label{black_box_watermarking}
		\end{minipage}
	}
	\caption{White-box (a) vs black-box (b) DNN watermark recovery.}
	\label{white&black_box}
\end{figure}

Based on the data accessible to the watermark extractor, we can divide DNN watermarking techniques into white-box and black-box techniques. As depicted in Figure \ref{white_box_watermarking}, if the internal parameters of the DNN models are available, we say that watermark recovery  is carried out in a white-box modality. The internal parameters may correspond directly to the model weights, or to the activation of the neurons in correspondence to specific inputs. In the case of multi-bit watermarking, the extractor, now called watermark decoder, extracts the concrete message bits the watermark consists of. For zero-bit watermarking, the watermark detector, must decide about the existence of a specific watermark.  In some algorithms (e.g. ~\cite{uchida2017embedding, Nagai_2018,chen2019deepmarks, rouhani2019deepsigns}), the key $K$ is  chosen independently from the original loss function $E_{0}$ and the internal parameters $\theta$, during a key generation step.

With black-box watermarking (see Figure \ref{black_box_watermarking}), only the final output of the DNN is accessible. In this case, the watermark is recovered by querying the model and checking the output of the DNN in correspondence to a set of properly chosen inputs. During the entire decoding or detection process, the architecture and the internal parameters of the DNN model are totally blind to the decoder or detector. In other words, the only thing that can be controlled are the inputs used for querying the network and the outputs corresponding to the queries. Thus, the main target of black-box watermarking is to train the DNN model in such a way that it outputs particular labels $z_{i}$ for certain inputs $x_{i}$'s. The inputs $x_{i}'s$ may be secret or not. In the former case they play a role similar to a decoding/detection key.

%Thus, given original loss function $E_{0}$, the Mark step inputted initial parameters $\theta_{ori}$ and $wk$ outputs trained parameters $\theta_{tra}$ and the trained model $F^{\prime}_{\theta_{tra}}$. The Verify step is a deterministic polynomial-time algorithm that given model parameters $\theta$ (or feeding selected images to the model to get parameters $\theta$ as a feedback) and (secret) extraction key $vk$ outputs extracted watermark message $wk$. \MB{First, the distinction here is not based on the watermark embedding, so you should focus on watermark extraction only. Second you consider only static watermarking, while the white box watermarking can refer to both dynamic and static watermarking.}

\subsubsection{Static vs. dynamic DNN watermarking}
\label{classification of static and dynamic}

\begin{figure}[htpb]
	\centering
	\subfloat[]{
		\begin{minipage}[t]{1\linewidth}
			\centering
			\includegraphics[scale=.32]{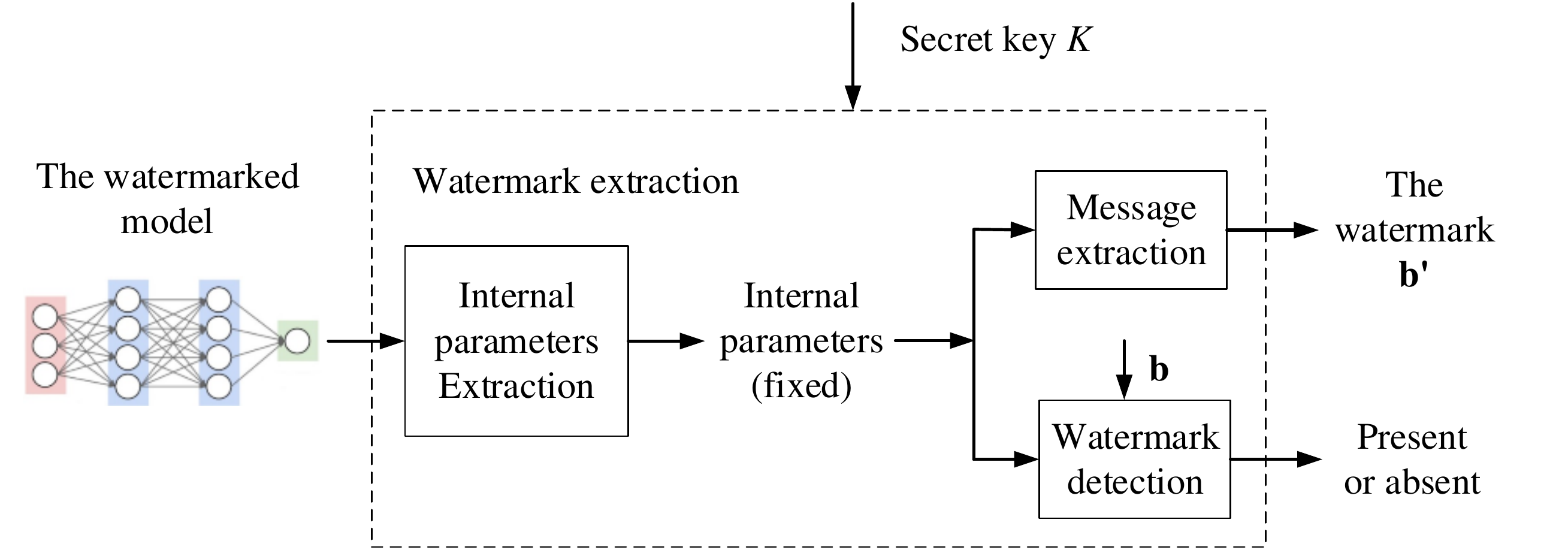}
			\label{static_watermarking}
		\end{minipage}
	} \\
	\subfloat[]{
		\begin{minipage}[t]{1\linewidth}
			\centering
			\includegraphics[scale=.33]{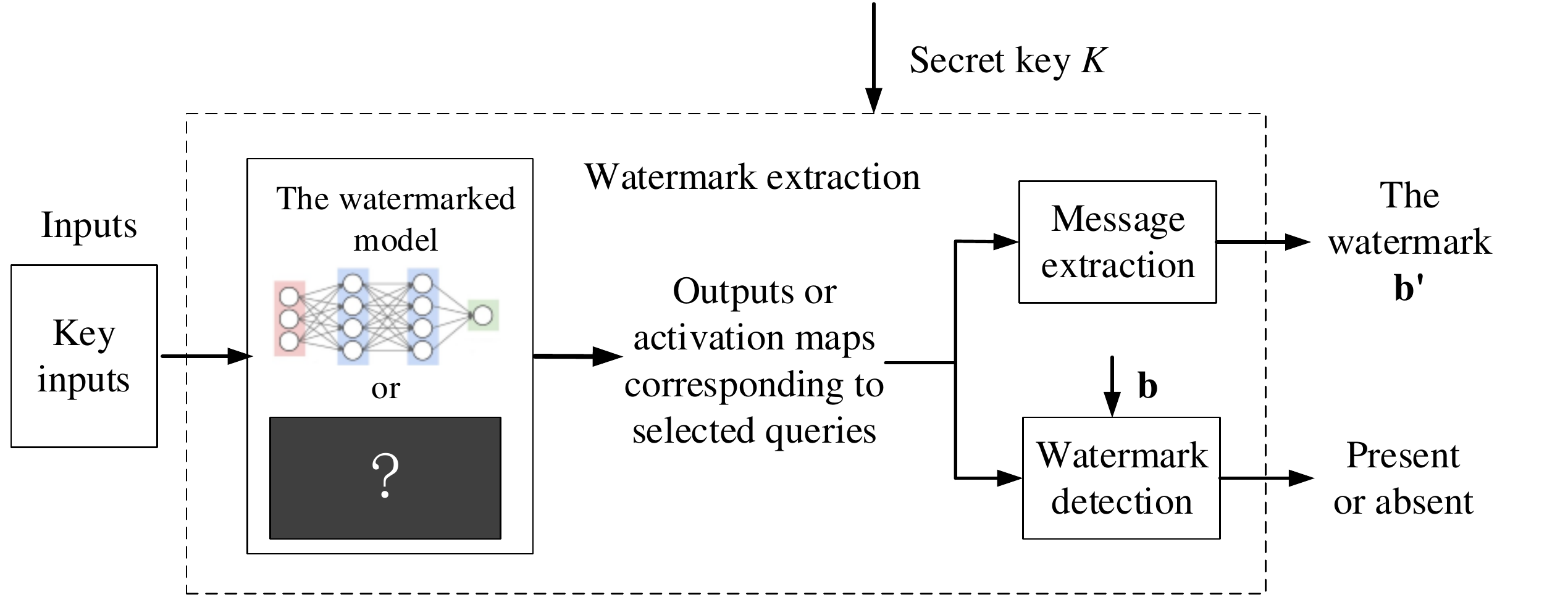}
			\label{dynamic_watermarking}
		\end{minipage}
	}
	\caption{Static (a) vs Dynamic (b) DNN watermarking}
	\label{static&dynamic}
\end{figure}

A very important distinction of DNN watermarking techniques leads to the definition of static and dynamic watermarking. Static DNN watermarking methods, like those described in \cite{uchida2017embedding, Nagai_2018, chen2019deepmarks}, embed the watermark into the weights of the DNN model. Such weights are determined during the training phase and assume fixed values that do not depend on the input of the network. With dynamic watermarking, instead, the watermark is associated to the behaviour of the network in correspondence to specific inputs, sometimes called triggering inputs, or key inputs (see for instance \cite{rouhani2019deepsigns, le2019adversarial, zhang2018protecting}). Even in this case, the watermark is embedded by properly chosen the network weights, however the watermark is retrieved indirectly by looking at the impact of the watermark on the behaviour of the network. If the watermark is recovered by looking at the final output of the model (sometimes referred to as {\em DNN-output} watermarking), the watermark can be extracted in a black-box manner, since access to the internal status of the network is not required. When the watermark is associated to the activation map of the neurons in correspondence to certain inputs, as in \cite{rouhani2019deepsigns}, white-box extraction is required. The distinction between static and dynamic watermarking is illustrated in Figure \ref{static&dynamic}.

%As shown in Fig. \ref{static_watermarking}, we recognize the kind of methods using the fixed parameters as the host for watermark as static watermarking. Inversely, when the host for a DNN watermarking algorithm is variable with the inputs (input-dependent) as in Figure \ref{dynamic_watermarking}, we define this algorithm as dynamic watermarking algorithm. For instance, weights are fixed parameters after training, no matter what the inputs of DNN models are \cite{uchida2017embedding, Nagai_2018, chen2019deepmarks}. Whereas, values of activation map hinge entirely on the inputs of DNN models \cite{rouhani2019deepsigns}. As for some classification-based methods(\cite{le2019adversarial, zhang2018protecting}), even if the internal parameters are not necessary during watermark detection, the rationale underneath is employing the character of input-dependent parameters to output particular labels when fed certain inputs. Thus, those DNN watermarking algorithms that are based on activation maps or classifications apparently are dynamic watermarking.

\subsection{Relationship among different DNN watermarking classes}

Table \ref{taxonomy} summarises the various perspectives that can be adopted to classify watermarking techniques and their applicability to conventional (multimedia) and DNN watermarking. The symbol $\surd$ indicates that the corresponding classification criterion is suitable for a specific watermarking technique, while \ding{55} represents inappropriateness.

Analyzing the intersection among these classifications, we see that dynamic watermarking can be implemented both in white box (activation-map based watermarking) and in black-box modality (DNN-output watermarking). On the contrary, static watermarking can only be achieved in a white-box mode, since it requires the access to the internal parameters of the network. In other words, white box can be static and dynamic, but black-box can only be dynamic. Referring to prior researches, white-box methods have been used for both multi-bit, and zero-bit watermarking, while the black-box modality, due to its lower expressivity, is more often used in conjunction with zero-bit watermarking.

\begin{table}
	\caption{Taxonomy of classic and DNN watermarking.}
	\label{taxonomy}
	\centering
	\setlength{\tabcolsep}{1mm}{
		\begin{tabular} {p{3em}p{6em}p{6em}p{6em}}
			\toprule
			& Multi/Zero-bit & Robust/Fragile & Blind/Non-blind\\
			\midrule
			Classic & $\surd $ & $\surd$ &$\surd$ \\
			DNN  & $\surd $ & Robust only & \ding{55} \\
			\midrule
			\midrule
			& Informed/Non-informed & White/Black-box & Static/Dynamic\\
			\midrule
			Classic &$\surd$  & \ding{55} & \ding{55}\\
			DNN  &  $\surd$ & $\surd$ &$\surd$\\
			\bottomrule
	\end{tabular}}
\end{table}

\section{Requirements}
\label{Sec:requirements}
In this section, we pause to discuss the main requirements that DNN algorithms must satisfy. While many of them are inherited by classical multimedia watermarking, there are some important differences that are worth attention. The most common requirements for DNN watermarking are summarised in Table \ref{tab:table}, and briefly discussed in the following.

\begin{table*}
\caption{List of requirements for DNN watermarking.}
	\begin{tabular}{m{2cm}<{\raggedright}m{14.5cm}<{\raggedright}}
		\toprule
		\textbf{Requirement}     & \textbf{Description}      \\
		\midrule
		Robustness & The embedded watermark should resist different kinds of processing.\\ \hline
		Security & The watermark should be secure against intentional attacks from an unauthorized party.\\ \hline
		Fidelity & The watermark embedding should not significantly affect the accuracy of the target DNN architectures.      \\ \hline
		Capacity & A multi-bit watermarking scheme should allow to embed as much information as possible into the host target DNN.\\ \hline
		Integrity & The bit error rate should be zero (or negligible) for multibit watermarking and the false alarm and missed detection probabilities should be small for the zero-bit case.\\ \hline
		Generality     & The watermarking methodology can be applied to various DNN architectures and datasets.     \\ \hline
		Efficiency & The computational overhead of watermark embedding and extraction processes should be negligible.\\
		\bottomrule
	\end{tabular}
\label{tab:table}
\end{table*}

\subsection{Robustness}

Together with security, robustness is one of the three corners of the watermarking trade-off triangle. It refers to the possibility of recovering the watermark from a perturbed version of the host content, be it a multimedia document or a DNN. Usually, it is required that the watermark survives at least the most common manipulations that the host content may undergo in its lifecycle. For DNNs, the two most common operations the watermark should be robust to are: fine-tuning and network pruning.

\textbf{Fine-tuning.} Fine tuning is a common operation related to transfer learning. It consists in retraining a model that was initially trained to solve a given task, so to adapt it to solve a new task (possibly related to the original one).
Computationally, fine-tuning is by far less expensive than training a model from scratch, hence it is often applied by users to adapt a pre-trained model to their needs. Needless to say, when sufficient training data is not available for the user, model fine-tuning can help avoiding over-fitting problems, at least to a certain extent. Since fine-tuning alters the weights of the watermarked model, it is necessary to make sure that the watermark is robust against a moderate amount of fine-tuning.

\textbf{Network pruning.} This is a common strategy to simplify a complex DNN model to deploy it into low power or computationally weak devices like embedded systems or mobile devices. During pruning, the model weights whose absolute value is smaller than a threshold are cut-off to zero. The threshold is set in such a way that the accuracy of the model does not decrease significantly. Of course, network pruning changes the internal parameters of the model, hence it is necessary to make sure that the embedded watermark is resistant to this operation.

\subsection{Security}
\label{subsec.sec}

This criteria deals with malicious manipulations, by assuming that an attacker, partially or fully aware of the watermarking algorithm, is present and attempts to destroy it, with the further goal of using the watermarked content illegally. For a secure DNN watermarking algorithm, the loss of the watermark should be obtainable only at the expense of a significant degradation on the quality of the host model. Until now, research has focused mainly on two kinds of intentional attacks, watermark overwriting and surrogate model attack.

\textbf{Watermark overwriting}.
A third-party who is aware of the methodology used for DNN watermarking (but does not know the private key used to embed the watermark) may try to embed a new watermark in the model and overwrite the original one. An overwriting attack, then, inserts an additional watermark into the model to make the original watermark undetectable. Several degrees of security can be defined according to the knowledge the attacker has about the watermark. When the watermark is inserted directly in the network weights, for instance, more powerful attacks can be conceived if the attacker knows the exact layers the watermark is embedded in.

\textbf{Surrogate model attack}.
In a surrogate model attack, the adversary tries to replicate the functionality of a target DNN, by feeding it with a series of requests and using the output provided by the network to train a surrogate model, which mimics the original network. During the attack, the adversary has no knowledge on the exact architecture of the model and limited access to the original training dataset. Ideally, if the network targeted by the attack is watermarked, the surrogate network trained by querying the watermarked model should also inherit the watermark, thus making it possible to recover the watermark from the surrogate model.

\subsection{Fidelity}
Fidelity represents the second corner of the watermarking trade-off triangle. It basically requires that the presence of the watermark does not degrade the {\em quality} of the watermarked object. In the case of DNNs, this means that the watermark should have a limited impact on the performance of the watermarked model $F_{\theta^\prime}$.
More specifically, the watermarked model $F_{\theta'}$ should guarantee a performance level that is as close as possible to that achieved by a model $F_\theta$ trained on the same task without caring about the watermark. As we will see later on, in some cases, training the network by adding a watermark embedding term to the loss function can even be beneficial, since it reduces the risk of overfitting. In the following sections, fidelity will be measured in terms of the Test Error Rate (TER) achieved on the task the model is designed for.

\subsection{Capacity}
Capacity is the third corner of the watermarking trade-off triangle. It is defined as the number of bits the watermark message consists of. As such, it can be referred only to multi-bit watermarking, since in the zero-bit case, the watermark does not convey any payload. While a large payload is a desirable feature of a watermarking algorithm, increasing the payload conflicts directly with the robustness requirement. The most straightforward way to increase the robustness, while also observing the fidelity criterion, in fact, is to spread the watermark over a large number of host samples (the network weights in the DNN case), thus rapidly exhausting the room available for watermark embedding. As an alternative, error correction coding can be used on top of the watermarking algorithm, to allow the recovery of some bits possibly lost because of the manipulation of the DNN. Even in this case, however, the room available to host the net payload decreases, thus raising again the necessity to find a tradeoff between the number of redundancy bits, which in turn determines the error correcting capability of the code, and the actual payload of the watermark.

\subsection{Integrity}
In the absence of model modifications, the extracted watermark $\mathbf{b}^\prime$ should be equal to $\mathbf{b}$. We use the Bit Error Rate (BER) to evaluate the integrity of a multi-bit algorithm. Ideally, in the absence of processing or attacks, the BER should be equal to zero. Classically, this is a characteristic achievable by the class of informed watermarking algorithms \cite{barni2004watermarking}, while it can not be guaranteed by non-informed spread spectrum techniques \cite{cox1997secure}. Interestingly, several DNN watermarking algorithms directly generate the model weights in such a way that the watermark can be extracted without errors. This resembles very closely the informed embedding paradigm \cite{miller2004applying}, according to which embedding is achieved by applying a signal-dependent perturbation to the to-be-watermarked sequence. In this way, the embedding procedure results in a watermark which, in the absence of modifications, can be recovered with no errors.

The integrity requirement assumes a different meaning in the case of zero-bit watermarking. In this case, watermark recovery can be stated as a detection problem and as such is prone to two kinds of errors: false detection and missed detection\footnote{The terms false positive and false negative are also used.}. The former refers to the probability that the presence of the watermark is a detected in a non-watermarked model, while the second indicates the probability that the watermark can not be recovered from a watermarked network. As it is known from statistical detection theory \cite{kay1993fundamentals}, decreasing the false detection probability comes at the price of an increase of the missed detection probability (and viceversa), so a trade off must be looked for. In DNN watermarking, it is pretty easy to design the watermark embedding algorithm in such a way that the missed detection probability is equal to zero\footnote{Once again this possibility is due to the adoption of an informed embedding strategy.}, while it is impossible to guarantee that the false detection probability is equal to zero. The only possibility, then, is to carry out a statistical analysis to give an estimate of the false detection probability and design the watermarking algorithm so that such a probability is acceptably small.

\subsection{Other Watermark Requirements}
Based on the choice of a specific watermarking modality among those listed in Section \ref{subsec: DNN model}, and on the envisaged application, other criteria may need to be satisfied.

\emph{Generality:} To be effective, a DNN watermarking algorithm should be applicable to a wide variety of architectures carrying out different tasks. In this sense, the generality of an algorithm refers to its applicability to architectures and models other than those on which the algorithm was initially tested and developed. This also implies that the performance of the algorithm do not depend too heavily on the to-be-marked model.

\emph{Efficiency:} Efficiency refers to the computational overhead needed to train the DNN by simultaneously teaching the model to carry out the the target task and embedding the watermark. Training a DNN is always an expensive process, embedding the watermark into the DNN should not add an unaffordable extra burden to it.

\section{Static Watermarking Algorithms}
\label{sec:static_algorithm}

In this section, we describe some examples of static watermarking algorithms. The goal is to illustrate with practical examples, how the watermark can be embedded directly into the weights of the network model and later on extracted from them. An exhaustive list of static DNN watermarking algorithms, is provided at the end of the section. 

So far, we have discussed the main properties of DNN watermarking by referring to generic models with generic inputs. By considering that the great majority of the watermarking algorithms proposed until now have been applied to image classification networks, in the following (Sections \ref{sec:static_algorithm}, \ref{sec: dynamic_algorithm} and \ref{sec.attacks}) we will always assume that the inputs of the DNN are images which must be classified into one of several possible classes.~\\

\subsection{Uchida et al. 's algorithm \cite{uchida2017embedding}} This is one of the first multi-bit techniques embedding the watermark message directly into the weights of the network model.

For a selected convolutional layer, let $(s, s)$, $d$, and $n$ represent, respectively, the kernel size of the filters, the depth of the input and the number of filters. Ignoring the bias term, the weights of the selected layer can be denoted by a tensor $ \mathbf{W} \in \mathbb{R}^{s\times s\times d\times n}$.
Before embedding the watermark into the weights, the tensor $\mathbf{W}$ is {\em flattened} according to the following steps: i)  calculate the mean of $\mathbf{W}$ over the $n$ filters, getting $\overline{\mathbf{W}} \in \mathbb{R}^{s\times s\times d}$ with $\overline{W}_{ijk} = \frac{1}{n} \sum_{h=1}^{n}W_{ijkh} $, in order to eliminate the effect of the order of the filters; ii) flatten $\overline{\mathbf{W}}$ producing a vector $\mathbf{w} \in \mathbb{R}^{v}$ with  $v=s\times s\times d$.
Watermark embedding is achieved by training the DNN with an additional loss term ensuring that the watermark bits can be correctly extracted from ${\bf w}$.
Specifically, the loss function $E(\mathbf{w})$ used to train the network and simultaneously embed the watermark into $\mathbf{w}$ is given by:
\begin{equation}\label{lossfuction}
	E(\mathbf{w}) = E_{0}(\mathbf{w}) + \lambda E_{R}(\mathbf{w})
\end{equation}
where $E_{0}(\mathbf{w})$ represents the original loss function of the network (ensuring a good behavior with regard to the classification task), $E_{R}(\mathbf{w})$ is the regularization term added to ensure correct watermark decoding, and $\lambda$ is a parameter adjusting the tradeoff between the original loss  term and the regularization  term. As we said, the goal of $E_{R}(\mathbf{w})$ is to make sure that the watermark is extracted from $\mathbf{w}$ with no errors.
Specifically, $E_{R}(\mathbf{w})$ is given by
\begin{equation}\label{regulization}
	E_{R}(\mathbf{w}) = -\sum_{j=1}^{l}(b_{j}\log(y_{j})+(1-b_{j})\log(1-y_{j}))
\end{equation}
\noindent where $b_{j}$ is the $j-$th bit of the watermark message (whose length is equal to $l$), and $y_j$ is the corresponding bit extracted by the watermark decoder. In particular, we have:
\begin{equation}% \label{regulization}
	y_{j}=\sigma\bigg(\sum_{i}X_{ji}w_{i}\bigg)
\end{equation}
where $\sigma(\cdot)$ is the sigmoid function defined as\footnote{We introduced the parameter $\gamma$ for sake of generality; in \cite{uchida2017embedding}, we have $\gamma = 1$.}:
\begin{equation} \label{sigmoid}
	\sigma(x)=\frac{1}{1+e^{-\gamma x}}.
\end{equation}
Here $\bf{X}$ is a spreading matrix, somewhat playing the role of a watermarking key, whose $j-$th row ($X_j$) is responsible of spreading the $j-$th bit onto a pseudorandom direction. In particular, in \cite{uchida2017embedding}, $\bf{X}$ is built by randomly drawing its elements according to a zero mean, unitary variance, Gaussian distribution $N(0,1)$. We observe that with this choice all the bits in $\bf{b}$ are cast into the same weights, however, they can be recovered with no error due to the statistical orthogonality of the rows of $\bf{X}$.

Watermark retrieval is pretty simple, since it consists in computing the projection of $\mathbf{w}$ onto each $X_j$, and thresholding the projection at 0, that is:
\begin{equation}
	\hat{b}_j =\begin{cases}
		1 & \sum_{i}X_{ji}w_{i} \geq 0,\\
		0 & \text{otherwise}.
	\end{cases}
\end{equation}

By training a wide residual network (WRN \cite{Zagoruyko_2016}) on the CIFAR-10 dataset (whose accuracy when trained without watermark corresponds to a test error rate equal to 8.04\%), Uchida et al's demonstrated that watermark embedding can be achieved without impacting significantly the classification accuracy, as reported in Table \ref{experiments_of_Uchida}. Interestingly, due to the use of the spreading matrix $\mathbf{X}$, it is possible to embed a payload which is larger than the number of weights available for embedding. The number of weights for the convolutional levels 2, 3 and 4, in fact, are 576, 1152, and 2304, however the maximum embeddable payload with BER = 0 is, respectively, 1024, 2048 and 4096. As it can be seen from the table, for a given convolutional layer, the payload is limited by the \emph{Integrity} requirement. When the payload exceeds a certain level, in fact, it is no more possible to embed the watermark ensuring a zero BER.

The robustness of Uchida et al's watermarking algorithm is exemplified in Tables \ref{Uchida_fine_tuning} and \ref{uchida_pruning}. With regard to fine tuning (Table \ref{Uchida_fine_tuning}), the watermarked model was fine-tuned for 20 epochs on the same CIFAR-10 dataset. As it can be seen, the test error does not change significantly, and the watermark BER remains zero. Table \ref{uchida_pruning}, reports robustness results against pruning. Specifically, we randomly set to zero a percentage p\% of the trainable parameters of the embedding layer. The results in the table prove the good robustness of Uchida et al's algorithm against pruning. On the other hand, overwriting the watermark with additional 256-bits (using a different spreading matrix) resulted in a large BER, thus showing the weakness of the algorithm against an intentional overwriting attack. 

\begin{table}
	\caption{TER and BER for different convolutional layers with various payloads, for Uchida et al.'s algorithm \cite{uchida2017embedding}.}
	\centering
	\setlength{\tabcolsep}{0.3mm}{
		\begin{tabular}{c|cc|cc|cc}
			\hline
			Payloads& \multicolumn{2}{c|}{Conv 2}& \multicolumn{2}{c|}{Conv 3}& \multicolumn{2}{c}{Conv 4}\\
			\cline{2-7}
			(bit) & TER(\%) & BER(\%)  & TER(\%) & BER(\%)& TER(\%) & BER(\%)\\
			\hline
			256& 7.97&0& 7.98& 0&7.92&0\\
			512& 8.47&0&8.22&0& 7.84&0\\
			1024& 8.43&0& 8.12 &0& 7.84&0\\
			2048&8.36 & 28.34  & 8.93 &0&7.75&0\\
			4096&8.25& 29.12&8.46&26.17&8.60&0\\
			\hline
	\end{tabular}}
	\label{experiments_of_Uchida}
\end{table}

\begin{table}
	\caption{TER and BER for robustness against fine-tuning for Uchida et al.'s algorithm \cite{uchida2017embedding}.} \label{Uchida_fine_tuning}
	\setlength{\tabcolsep}{1mm}{
		\begin{tabular}{c|c|ccc}
			\hline
			Embedded  & Payload &  TER & TER & BER\\
			layer&  (bit) & (\%) & after attack(\%)&  after attack (\%) \\
			\hline
			Conv 2 & 256  & 7.97 & 8.11 & 0\\
			\hline
			\multirow{2}{*}{Conv 3}& 256  & 7.98 & 7.58 & 0\\
			& 1024 &8.12& 8.17&0 \\
			\hline
			\multirow{2}{*}{Conv 4}& 256  & 7.75 & 7.90 & 0\\
			& 4096  & 8.60 & 8.26 & 0\\
			\hline
	\end{tabular}}
\end{table}
\begin{table}
\caption{TER and BER for robustness against parameter pruning for Uchida et al.'s algorithm \cite{uchida2017embedding}.}\label{uchida_pruning}
\setlength{\tabcolsep}{1mm}{
	\begin{tabular}{c|c|c|ccc}
		\hline
		Embedded &Payload  &$p$ &TER & TER after & BER after\\
		layer&(bit) &  &(\%) & attack (\%)&  attack (\%)\\
		\hline
		Conv 2 & 256 &10\% & 7.97 & 10.23 & 0\\
		\hline
		\multirow{2}{*}{Conv 3}& 256 &10\%  & 7.98 & 8.49 & 0\\
		& 512 &10\% &8.22&8.80 &0 \\
		\hline
		\multirow{2}{*}{Conv 4}& 256 & 20\%  & 7.75 & 8.73 & 0\\
		& 4096 &10\% & 8.60 & 9.69 & 0\\
		\hline
	\end{tabular}}
\end{table}

\subsection{ST-DM DNN watermarking~\cite{li2020spread} }
Uchida's algorithm is designed according to a classical spread spectrum strategy \cite{cox1997secure}. Given that the weights are generated directly in such a way that the watermark is recovered correctly, the embedding procedure obeys the informed embedding paradigm \cite{miller2004applying}. In contrast, no attempt is made to exploit informed coding to increase the payload or diminish the impact of watermarking on the accuracy of the watermarked model. An example of the use of informed coding for DNN watermarking is given in \cite{li2020spread}. The algorithm proposed in such a work relies on a new loss function defined as follows:
\begin{equation}\label{loss_function2}
	E_{\text{ST-DM}}(\mathbf {w})=E_{0}(\mathbf{w})+\lambda E_{\text{ST-DM}}(\mathbf{w})
\end{equation}
with the new regularization term $E_{\text{ST-DM}}$ given by:
\begin{equation}\label{ours_regularization}
	E_{\text{ST-DM}}(\mathbf{w}) = -\sum_{j=1}^{l}(b_{j}\log(y_{j})+(1-b_{j})\log(1-y_{j})),
\end{equation}
where $\mathbf{y}$ is obtained by approximating the decoding function employed by ST-DM.
To see how, let us consider two uniform interleaved quantizers $\QQ_0$ and $\QQ_1$ associated, respectively, to bit 0 and 1, and defined by the following codebooks
\begin{align}
	&\UU_0 = \left\{ k\Delta, k \in \mathbb{Z} \right\}\\
	&\UU_1 = \left\{ k\Delta + \Delta/2, k \in \mathbb{Z}
	\right\}.
	\label{eq.codebook_DM}
\end{align}
where $\Delta$ is the quantization step. Given a watermarked sample $w_m$, Dither-Modulation (DM) decoding works by looking for the entry in $\UU_0 \cup \UU_1$ closest to $w_m$ and see if such an entry belongs to $\UU_0$ or $\UU_1$, which is equivalent to applying the following decoding function:
\begin{equation}
	\phi_{DM}(w) = \arg \min_{b=0,1} (\min_{u_{k}\in {\mathcal U}_b} | w_m - u_{k}|).
\end{equation}
DM watermarking follows directly from the above formulation, and is achieved by quantizing the host sample $w$ either with $\QQ_0$ (to embed a 0 bit) or $\QQ_1$ (to embed a 1 bit). ST-DM watermarking is achieved by applying DN watermarking to the projection of a sequence of host samples on a spreading direction.
	
In \cite{li2020spread}, watermark decoding is achieved by applying $\phi_{DM}$ to the projection of $\mathbf w$ (defined as in the previous section) over the directions determined by the rows of the pseudorandom matrix $\mathbf{X}$, that is:
\begin{equation}
	y_j = \phi_{DM}\bigg(\sum_i X_{ji} w_i\bigg).
\end{equation}
Due to the non-continuous nature of $\phi_{DM}$, the direct use of $\phi_{DM}$ in Eq. \eqref{ours_regularization} would make it difficult the application of back-propagation to train the network, hence $\phi_{DM}$ is approximated with a smooth function $\theta()$, defined as:
\begin{equation}\label{ours_function}
	\theta(x) = \frac{e^{\alpha \sin \beta x}}{1 + e^{\alpha \sin \beta x}},
\end{equation}
\begin{figure}
	\centering
	\includegraphics[scale=.40]{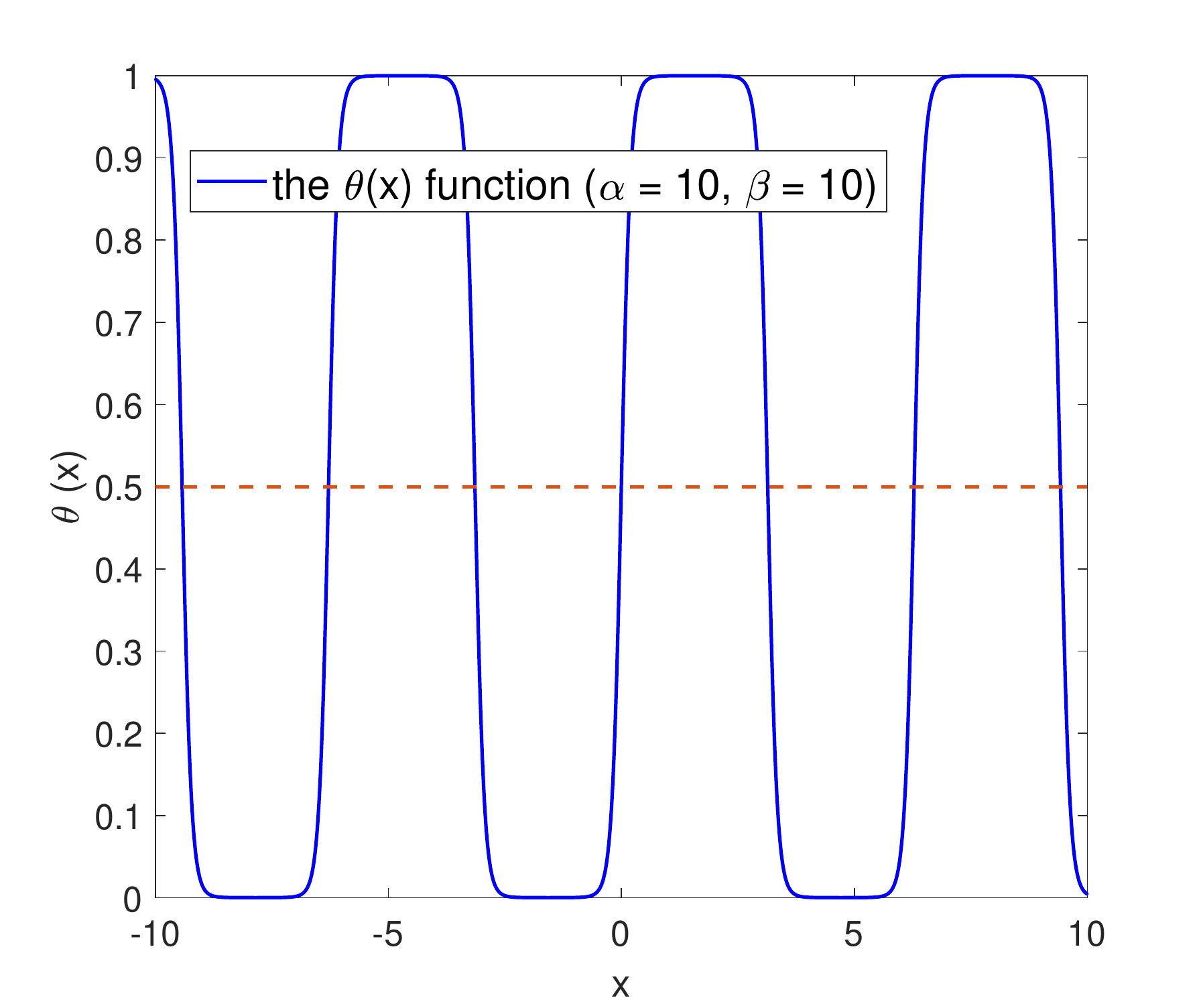}
	\caption{Behavior of the function $\theta(x)$ in \cite{li2020spread}.}
	\label{theta_function}
\end{figure}
where $\alpha$ determines the smoothness of $\theta$ and $\beta$ its periodicity, directly related to the quantization step defining the ST-DM watermarking algorithm. An example of the function $\theta(x)$ for $\alpha = \beta = 10$ is shown in Fig. \ref{theta_function}. Finally, the $y_j$'s in  Eq. \eqref{ours_regularization} are computed as $y_j = \theta\left(\sum_i X_{ji} w_i\right)$, and watermark extraction is carried out as follows:
\begin{equation}
	\hat{b}_j =\begin{cases}
		1 & \theta\left(\sum_{i}X_{ji}w_{i}\right) \geq 0.5,\\
		0 & \text{otherwise}.
	\end{cases}
\end{equation}
	
The results of some experiments showcasing the performance of ST-DM DNN watermarking are reported in Tables \ref{experiments_of_us} and \ref{ours_fine_tuning} obtained by using the same setting as in Tables \ref{Uchida_fine_tuning} and \ref{uchida_pruning}, with $\alpha=\beta=10$. Generally speaking, ST-DM watermarking allows a larger capacity than conventional spread spectrum, with a similar level of robustness.
	
\begin{table}
	\caption{TER and BER for different convolutional layers with various payloads for Li et al.'s algorithm \cite{li2020spread}.}
	\centering
	\setlength{\tabcolsep}{0.3mm}{
	\begin{tabular}{c|cc|cc|cc}
			\hline
			Payloads& \multicolumn{2}{c|}{Conv 2}& \multicolumn{2}{c|}{Conv 3}& \multicolumn{2}{c}{Conv 4}\\
			\cline{2-7}
			(bit) & TER(\%) & BER(\%)  & TER(\%) & BER(\%)& TER(\%) & BER(\%)\\
			\hline
			256& 8.20 & 0& 8.15 & 0&8.20 & 0 \\
			512& 7.75 &0&8.07 &0& 8.57 &0\\
			1024& 7.63 &0& 8.12 &0& 7.84&0\\
			1200& 7.96 &0 & 8.31 &0  & 8.06  &0\\
			2048&8.36 &5&  7.85 &0 &8.03&0\\
			2400& 8.82& 15.08& 8.22 & 0& 8.31 &0\\
			4096&8.35 &23.14 &8.35 &13.84 &7.64&0\\
			4800  &8.29& 24.33&8.26&14.06  & 8.65 & 0    \\
			\hline
	\end{tabular}}
	\label{experiments_of_us}
\end{table}
	
\begin{table}
	\caption{TER and BER for robustness against fine-tuning for Li et al.'s algorithm \cite{li2020spread}.} \label{ours_fine_tuning}
	\setlength{\tabcolsep}{1mm}{
	\begin{tabular}{c|c|ccc}
		\hline
		Embedded  & Payload &  TER & TER & BER\\
		layer&  (bit) & (\%) & after attack(\%)&  after attack (\%) \\
		\hline
		Conv 2 & 256  & 8.20& 8.02& 0\\
		\hline
		\multirow{2}{*}{Conv 3}& 256  & 8.15& 8.05 & 0\\
		& 1024 &7.79 &7.65&0 \\
		\hline
		\multirow{2}{*}{Conv 4}& 256  & 8.20& 7.93 & 0\\
		& 4096  &7.64& 7.43 & 0\\
		\hline
	\end{tabular}}
\end{table}
	
\subsection{DeepMarks.} 
In \cite{chen2019deepmarks}, Chen et al. exploit Uchida's algorithm to implement a traitor tracing watermarking system with anti-collusion capabilities. In a traitor tracing application, a different code is assigned to each different user of the network. Then, a watermark bearing the user's code is embedded within the network prior to its release to the user. If a non-authorized copy of the model is found, the content of the watermark allows to identify the user who illegally distributed it. To avoid that a subset of the users form a coalition, to produce a new model where all the watermarks are {\em mixed} together hence making it impossible to identify the users who redistributed the model illegally, Anti Collusion Codes (ACC) may be used to identify the users \cite{trappe2003anti}.
	%watermark embedding and extraction are carried out as in the original Uchida's algorithm.
	
In DeepMarks, the watermark bits are generated by using a ($v, k, 1$)-Balanced Incomplete Block Design (BIBD) code, which can be represented by its corresponding incidence matrix $C_{v\times n}$ where $n = (v^2 -v)/(k^2-k)$  and each elements of $C_{v\times n}$:	\begin{equation}
		c_{ij} = \left\{ \begin{array}{ll}
			1& \text{if}~ i-{th} \text{ value occurs in}~ j-{th} \text{ block}\\
			0& \text{otherwise}.
		\end{array} \right.
\end{equation}
By letting the bit complement of the columns of the incidence matrix $C_{v\times n}$ represent the code-vectors, the resulting $(v, k, 1)$-BIBD code can identify coalitions of up to ($k-1$) members chosen among $n$ users \cite{trappe2003anti}.

\subsection{Tartaglione et al. \cite{Tartaglione2020delving}} 
As an example of a static zero-bit watermarking system, we describe the method recently proposed by Tartaglione et al. in \cite{Tartaglione2020delving}. By leveraging on the superior robustness of zero-bit watermarking systems, and by adopting a properly designed loss function explicitly thought to increase the robustness of the watermark, such a system achieves a remarkable robustness against fine tuning and weights quantisation.
	%\TODO{Read the paper and include a brief description by reporting also some results. Also include this method in Table \ref{static_watermarking_summary}}
	
In contrast to the schemes described so far, in this case the watermarked coefficients are set before the training procedure starts and are not modified during training. To make the watermarked weights undistinguishable from the others (for security reasons), and by assuming that the weights of the to-be-watermarked architectures are initialised following a gaussian distributions ($\mathbf{w} \sim \mathcal{N}(\mu,\sigma)$), the watermarked weights are generated as follows. Let $b_i \in [0,1]$ be the watermark coefficient associated to the $j$-th weight $w_j$ (the choice of the weights carrying the watermark is made based on a pseudorandom number generator distributing the watermark across the entire model), the watermarked weights are generated as:
\begin{equation}
		w_j = 2\sigma(b_i - 0.5) + \mu  \qquad \forall w_j \in \mathbf{w}_b,
		\label{TartEmbed}
\end{equation}
where $\mathbf{w}_b$ is a vector with all the weights associated to the watermark element $b$.
To make sure that the watermarked weights are not modified during the training process, the watermarked weights are labelled and left unchanged during the gradient-based update rule, that is:
\begin{equation}
		w_j = w_j-[1-\text{ind}(w_j)]\eta \frac{\partial L}{\partial w_j}
\end{equation}
where $\eta$ is the learning rate and $\text{ind(}w_j)$ is the indicator function identifying the watermarked weights:
\begin{equation}
		\text{ind}(w_j) = \begin{cases}
			1 & if\quad w_j\in \mathbf{w}_b\\
			0 & \text{otherwise}.
		\end{cases}
\end{equation}
Watermark detection is achieved by inverting equation of Eq. \eqref{TartEmbed}:
\begin{equation}
	\hat{b_i} = \frac{w_j-\mu}{2\sigma}+0.5
\end{equation}
	
Robustness against fine tuning is obtained by adding a term to the training loss function forcing the watermarked network to be highly sensitive to even small changes of the watermarked weights. In this way, during fine tuning, such weights are not modified (or modified only slightly) hence limiting the effect of fine tuning on the watermark.
To do so, $R$ perturbed versions of the DNN are generated. The weights of such networks are perturbed by adding to them a noise term generated according to a normal distribution:
\begin{equation}
	w_j^r = w_j + \text{ind}(w_j) \Delta w_j^r,
		\label{perturb}
\end{equation}
where $w_j^r$ ($r = 1 \dots R$) are the weights of the perturbed networks and $\Delta w_j^r$ is the noise term added to perturb them.
The gradient of the loss function of the perturbed networks with respect to the perturbed weights is then computed and averaged across all the networks:
\begin{equation}
	g_i = \frac{1}{R} \sum_{r=1}^{R}\frac{\partial L}{\partial w_j^r}.
	\label{gi}
\end{equation}
Finally, the to-be-watermarked network is trained in such a way to maximise the sensitivity of the model to the watermarked weights. Specifically, the update rule of the DNN weights is defined by:
\begin{equation}
		w_j = w_j-[1-\text{ind}(w_j)]\bigg[\eta \frac{\partial L}{\partial w_j}-\gamma g_j\bigg],
\end{equation}
where $\gamma > 0$ is a properly selected hyper-parameter.
	
An excerpt of the experimental results reported in \cite{Tartaglione2020delving} is given in Table \ref{static_zerobit}, where 0.4\% of the model weights are used for conveying the watermark. According to the results, the influence of the watermark on the TER is negligible. The Pearson correlation between the original and extracted watermark is used to measure the robustness against fine tuning. As reported in Table \ref{static_zerobit_robustness},  this method shows good robustness against fine-tuning.
\begin{table}
	\caption{Performances of the watermarking system proposed in \cite{Tartaglione2020delving}. $R=0$ refers to a network trained without enforcing watermark robustness.}
	\begin{tabular}{ccccc}
		\hline
		Dataset&Architecture&Epochs&R&TER(\%)\\
		\hline			\multirow{4}{*}{MNIST}&\multirow{4}{*}{LeNet5}&\multirow{4}{*}{100}&0&0.78\\
		&&&1&0.83\\
		&&&4&0.82\\
		&&&16&0.85\\
		\hline
		\multirow{3}{*}{CIFAR-10}&\multirow{3}{*}{ResNet-32}&\multirow{3}{*}{350}&0&7.14\\
		&&&2&7.08\\
		&&&4&7.29\\
		\hline
	\end{tabular}\label{static_zerobit}
\end{table}

\begin{table}
	\caption{Fine-tuning attack on LeNet5 trained with different $R$ values according to the method described in \cite{Tartaglione2020delving}.}
	\begin{tabular}{ccc}
		\hline
		$R$ values&Epochs&Person correlation(\%)\\
		\hline		
		\multirow{3}{*}{0}&10&99.986\\
		&30&99.951\\
		&50&99.900\\
		\hline
		\multirow{3}{*}{4}&10&99.999\\
		&30&99.998\\
		&50&99.997\\
		\hline
		\multirow{3}{*}{16}&10&99.999\\
		&30&99.999\\
		&50&99.998\\
		\hline
	\end{tabular}\label{static_zerobit_robustness}
\end{table}

A summary of the static watermarking algorithms described so far is given in Table~\ref{static_watermarking_summary}, including their classification according to the taxonomy introduced in Sect. \ref{sec: taxononmy} and their main properties.~\\
	
\subsection{Other works}
Many other static watermarking schemes have been developed starting from the basic scheme by Uchida et al. \cite{uchida2017embedding}. 

Wang et al. \cite{wang2020watermarking}, for instance, propose to embed the watermark into early converging weights by bringing in an additional independent neural network to embed and extract the watermark. 

In \cite{wang2019attacks}, the undetectability of Uchida et al's watermark is studied, showing that the existence of the watermark can be easily detected by analyzing the statistical distribution of the weights. A solution to solve this problem is proposed in \cite{wang2019robust}, by relying on a generative adversarial network architecture (GAN). The watermarked model acts as the generator, while the discriminator  is in charge of detecting the watermarked weighs. The discriminator provides its feedback to the generator, which in turns is encouraged to embed the watermark in such a way that the watermarked weights are  statistically similar to the non-watermarked ones. A further development of the scheme described in \cite{wang2019robust} (called RIGA - \textbf Robust wh\textbf Ite-box \textbf {GA}n) is presented in \cite{wang2019riga} together with a detailed theoretical and experimental analysis.

Another interesting improvement of \cite{uchida2017embedding} has been proposed in  \cite{cortinas2020adam}, based on the observation that the adoption of the Adam optimiser introduces a dramatic variation on the histogram distribution of the weights after watermarking, which can be easily detected by the adversaries. To solve this problem, the authors propose to use an orthonormal projection matrix, and to include in the projection also the gradients of the weights. Then the Adam optimiser is run on the projected weights using the projected gradients.
	
Kuribayashi et al. \cite{kuribayashi2020deepwatermark}, propose a method that embeds the watermark into the selected weights via fine-tuning. After sampling the weights from the fully-connected layer, the original weights are substituted by the watermarked sampled weights by means of a quantization-based method (such as DM-QIM). To limit the impact that the watermark has on the accuracy of the network, both the fully-connected layer and the convolutional blocks ahead of the watermarked fully-connected layer are fine-tuned.
	
The DeepMarks scheme has been extended in \cite{chen2019deepattest} for protecting the rights of devices providers, and in \cite{chen2020specmark} for IPR protection of Automatic Speech Recognition (ASR) systems.
	
\begin{table*}\centering
	\scriptsize
	\caption{Summary of the main static watermarking algorithms described in Sect. \ref{sec:static_algorithm}.}
	\setlength{\tabcolsep}{0.1mm}{
		\begin{tabular}{cm{2cm}<{\centering}m{2cm}<{\centering}m{3cm}<{\centering}m{3cm}<{\centering}m{2cm}<{\centering}m{3cm}<{\centering}}
			\hline
			Algorithm&White/ Black-box&Multi/ Zero-bit&Methodology&Key generation&Payload (bits)& Robustness and Security\\
			\hline
			Uchida et al. \cite{uchida2017embedding}&White-box&Multi-bit&A regularization term is added to the loss function to embed the watermark into the model weights according to the SS principle.& SS matrix drawn from a standard normal distribution. & conv2:1024; conv3:2048; conv4:4096 & Moderate robustness against fine-tuning and pruning\\
			\hline
			Li et al. \cite{li2020spread}&White-box&Multi-bit&As Uchida's scheme with ST-DM-like regularization term.& SS matrix drawn from a standard normal distribution.& conv2:1200; conv3:2400; conv4:4800&Moderate robustness against fine-tuning and pruning\\
			\hline
			DeepMarks \cite{chen2019deepmarks}& White-box & Multi-bit & As Uchida's scheme with anti-collusion codebooks & Normally distributed SS matrix, plus anti-collusion codebooks & Basically the same as \cite{uchida2017embedding} & Moderate robustness against fine-tuning and pruning, Collusion attack \\
			\hline
			Tartaglione et al. \cite{Tartaglione2020delving} & White-box & Zero-bit & The watermarked weights are frozen during the training procedure. The loss function is designed so to maximize the sensitivity of the network to changes of the watermarked weights. & A pseudorandom number generator is used to select the weights to be watermarked. & Zero-bit  & Good robustness against fine-tuning and weights quantization\\
			\hline				
		\end{tabular}}\label{static_watermarking_summary}
\end{table*}

\section{Dynamic Watermarking}\label{sec: dynamic_algorithm}
As we already noted in Section \ref{classification of static and dynamic}, with dynamic methods, the watermark is associated to the behaviour of the DNN in correspondence to a set of properly selected inputs. If the inputs activating the watermark are kept secret, their role is analogous to a watermarking key. For this reason, in the following we will refer to them as key-inputs, or key-images in the case of DNN dedicated to image analysis. Due to their capability to {\em trigger} the desired dynamic watermarking behaviour, the key-inputs are sometimes referred to as watermark triggers. The key-inputs and the corresponding labels form a so-called \emph{key input-label pair}.

The choice of the inputs activating the behaviour associated to the watermark is a crucial one.
An important distinction can be made between systems wherein the key inputs  are {\em entangled} with the task the network is intended to solve and {\em non-entangled} ones \cite{jia2020entangled}. In the former case, the key-inputs are chosen within the class of inputs the network has been designed to work on. In the latter case, they correspond to outlier inputs that are not expected to be fed to the model in normal operative conditions.
The use of entangled key inputs, forces the model to learn features which are jointly used to analyse both the normal and the key inputs. In this way, it is more difficult for an adversary to separate the watermark and the training task, hence resulting in a more robust watermark. On the contrary, in the non-entangled case, the impact of the watermark on model accuracy is lower, since the network may learn distinct features for watermark encoding and the primary task it is intended for, but, arguably, the resulting watermark will be weaker and easily removed by fine tuning.

Depending on the network layer whose behaviour is modified by the watermark presence, we can classify dynamic watermarking methods in two main classes, \emph{activation-map} watermarking and \emph{DNN-output} watermarking. In the former case, the behaviour induced by the watermark is observed at the intermediate layers of the network, in the latter, the watermark presence can be revealed by observing only the output of the network. Since the activation maps are internal parameters of the DNN models, activation-map methods belong to the class of white-box watermarking. On the contrary, DNN-output algorithms belong to the black-box category.

\subsection{White-box dynamic watermarking of activation maps: the DeepSigns algorithm \cite{rouhani2019deepsigns}}
\label{activation_based}
As an example of white-box dynamic watermarking based on activation maps, we describe the DeepSigns method proposed in \cite{rouhani2019deepsigns}. As a matter of fact, DeepSigns can be applied in both white-box and black-box setting, in this section we focus on the white box version.

In DeepSigns, the behaviour of the watermarked activation map is defined by forcing the distribution of the map to follow
a Gaussian Mixture Model (GMM) for which the mean values of the Gaussian probability density functions (pdfs) satisfy certain conditions which in turn determine the embedded bits. The desired behaviour is enforced by adding two additional terms to the loss function used for training:
\begin{equation}
	L_{DS} = L_0 + \lambda_1 L_1 + \lambda_2 L_2,
	\label{eq:DSloss}
\end{equation}
where $L_0$ is the original loss function used to train the non-watermarked models, and $\lambda_1$ and $\lambda_2$ are weights used to balance the importance of the two additional terms of the loss.

The goal of the term $L_1$ is to let the distribution of the activation map hosting the watermark to be as close as possible to the desired distribution. Different Gaussian models are enforced for inputs belonging to different input classes, so to make sure that the network retains its classification capabilities. To be specific the term $L_1$ is defined as:
\begin{equation} \label{DeepSigns_loss2}
	L_1 = \sum_{i\in T}||\mu_{l}^{i}-f^{i}_{l}(x, \theta)||^{2}_{2}-\sum_{i\in T, j\notin T}||\mu_{l}^{i}-\mu^{j}_{l}||^{2}_{2},
\end{equation}
where $f^{i}_{l}(x, \theta)$ is the activation map corresponding to the input sample $x$ belonging to class $i$ at the $l^{th}$layer, $T$ is the set of target Gaussian classes selected to carry the watermark, and $\mu_{l}^{i}$ denotes the mean value of the $i^{th}$ Gaussian distribution at layer $l$. In the following, we indicate with $s$ the number of Gaussian models (input classes) chosen to host the watermark. The goal of the second term in Eq. \eqref{DeepSigns_loss2} is to ensure that the Gaussian models associated to different classes are far apart, hence resulting in better classification performance.

To understand the form of the term $L_2$, let us consider the procedure whereby the watermark is read from the activation map. Let $b^{s\times N}$ be a matrix with the watermark bits (the $i^{th}$ row of the matrix contains the bits embedded in the pdf of class $i$) and let $\mu^{s\times M}_l$ represent the mean values of the corresponding Gaussian distributions. We first multiply  $\mu^{s\times M}_l$ by a spreading matrix $\bf A$ whose role is analogous to that of ${\bf X}$ in Uchida et al.'s algorithm, then we pass the results through a sigmoid function. The watermark bits are computed by comparing the output of the sigmoid function with a threshold equal to 0.5. In formulas:
\begin{equation} \label{DeepSigns_embed}
	G^{s\times N}_\sigma = \sigma(\mu_l^{s\times M} \cdot A^{M\times N})	\notag
\end{equation}
\begin{equation}
	{\hat b}^{s\times N} =\begin{cases}
		1 & G^{s\times N}_\sigma \geq 0.5,\\
		0 & \text{otherwise}.
	\end{cases}\\
\end{equation}
where $M$ is the size of the activation map in the selected layer, that is the number of nodes of the fully-connected layer and $N$ indicates the length of the watermark sequence embedded within each Gaussian model.

The loss term $L_2$ is built in such a way to guarantee that the bits extracted as detailed above correspond to the correct bits. Such a goal is reached by letting $L_2$ correspond to the cross-entropy between the desired bit matrix $b^{s\times N}$ and $G^{s\times N}$:
\begin{equation}
	L_2 = - \sum_{j=1}^{N}\sum_{k=1}^{s}(b^{kj}\text{ln}(G^{kj}_\sigma)+(1-b^{kj})\text{ln}(1-G^{kj}_\sigma)).
\end{equation}
The term $L_2$ is computed only on the key images, since the watermark can be read only when the network is fed with such images. In \cite{rouhani2019deepsigns}, the key images are chosen as a subset of the input training data belonging to the selected classes (entangled watermark triggers). For example, for a network trained on the MNIST dataset, the selected class to carry watermark corresponds to a specific digit, e.g. the zero digit. Thus, the key images are a subset of randomly chosen images of the zero-digit class.

The experimental results reported in \cite{rouhani2019deepsigns}, prove that the DeepSigns watermark provides good
performances with regard to fidelity and robustness. In contrast, the watermark payload is quite limited as shown in Table \ref{DeepSigns_white_results}. To ensure a zero BER, DeepSigns can allow up to 64 bits for MNIST and up to 128 bits for CIFAR 10-CNN and CIFAR 10-WRN.

\begin{table*}
\caption{Performance achieved by DeepSigns on different marked models and datasets \cite{rouhani2019deepsigns}. 64C3(1)indicates a convolutional layer with 64 output channels and $3\times3$ filters applied with a stride of 1, MP2(1) denotes a max-pooling layer over regions of size $2\times 2$ and stride of 1, and 512FC is a fully-connected layer with 512 output neurons. In all cases BER = 0.}
\setlength{\tabcolsep}{0.5mm}{
	\begin{tabular}{cm{5cm}<{\centering}ccm{1.5cm}<{\centering}c}
			\toprule
			DNN Model Type&DNN Model Architecture&Dataset&Baseline TER& TER of Marked Model&Payload \\
			\midrule
			MLP (Multi-Layer Perceptron) & 784-512FC-512FC-10FC& MINST& 1.46\%& 1.87\% &4\\
			\midrule
			CNN  & $3\times 32\times32$-32C3(1)-32C3(1)-MP2(1)-64C3(1)-64C3(1)-MP2(1)-512FC-10FC& CIFAR10& 21.53\% & 19.3\% & 4\\
			\midrule
			WideResNet & Please refer to \cite{Zagoruyko_2016}&CIFAR10& 8.58\%&7.98\%& 128\\
			\bottomrule
	\end{tabular}}\label{DeepSigns_white_results}
\end{table*}

\subsection{Black-box dynamic watermarking}

With black-box dynamic watermarking, the watermark is encoded in the relationship between a selected set of key inputs and the corresponding outputs. For such a relationship to be able to characterise the watermark and to avoid interfering with the intended behaviour of the model, it is necessary that the key inouts are carefully chosen.

\subsubsection{Watermarking and DNN backdoors} 
Generally speaking, teaching the network to behave in a {\em special} way in correspondence to a small set of selected inputs, while working as usual on the other inputs, is analogous to embedding a backdoor within the network, hence drawing a strong relationship between dynamic DNN watermarking and DNN backdoors \cite{adi2018turning}.
In a backdoor attack \cite{li2020backdoor}, the attacker corrupts the training phase to induce a classification error, or any other erroneous behaviour, at test time.
Test time errors, however, only occur in the presence of a triggering event corresponding to a properly crafted input. In this way, the backdoored network continues working as expected for regular inputs, and the malicious behaviour is activated only when the attacker decides to do so by feeding the network with a triggering input.
With black-box dynamic watermarking, the activation of the backdoor does not correspond to a misbehaviour of the network, on the contrary it is instrumental for revealing the presence of the watermark and read its content.

Black-box dynamic watermarking methods can be classified according to the way the key-images\footnote{In this section we always refer to key-images since most works proposed so far focus on the watermarking of DNN with images inputs.} (or trigger images if we adopt a backdoor perspective) are chosen.
In a first category of methods, the key images are chosen within a set of existing images, either belonging to the class of images the network is supposed to classify (entangled key-images) or not. In the following we will refer to this kind of key images as {\em natural key images}. By interpreting watermarking as backdoor injection, we can say that in this class of methods, the events triggering the watermark behaviour are the images themselves. In a second case, the key-images are handcrafted so to satisfy certain conditions. From a backdoor perspective, this corresponds to embed within the key images a specific triggering pattern, that may assume the form of a hidden signal or even a visible pattern.

In the following, we describe some different black-box watermarking methods characterised by different ways of choosing the key-images. With few exceptions (see \cite{chen2019blackmarks}), most black-box schemes belong to the class of zero-bit watermarking.

\subsubsection{Natural key images}

This branch of algorithms choose the key images without manipulating them. For this kind of methods, it's vital to find a proper way to identify a set of key images for which meeting the requirements listed in Section \ref{Sec:requirements} is not too difficult. Some example of methods belonging to this category among those proposed so far are explicit described in the following.~\\

\noindent\textbf{Yossi et al.~\cite{adi2018turning}}~were the first to suggest the use of backdoors for DNN watermarking. Specifically, a set of non-entangled key images is first chosen, then the labels associated to the key-images are sampled randomly from all the input classes excluding original labels predicted by the non-watermarked network. Two approaches are investigated for watermarking. According to the first approach, the network is first trained without key images, then a second phase of training is applied by using the key images as well. The second approach consists in training the network from scratch by also including the key images. To detect the presence of the watermark (the methods described in \cite{adi2018turning} is a zero-bit algorithm), the labels predicted on the key-images are compared with the desired one: for non watermarked models the agreement will generally be very low, while for the watermarked models most of the key-images will be assigned the correct label. The detection threshold is set based on the false positive probability estimated by modelling the probability that a non-watermarked network fed with a key image outputs the correct key label with a binomial distribution.

The validity of the system has been demonstrated by applying it to a ResNet-18 model trained on CIFAR10, CIFAR 100 and ImageNet datasets. 100 non-entangled key images with an abstract content have been used with key labels randomly chosen from the labels of the training set. The results show that the watermark obtained in this way  achieves a good robustness against a fine-tuning attack, as reported in Table \ref{backdoor_experiments}. For the fine-tuning experiments, the models were fine-tuned for 20 epochs on the on STL-10 datasets \cite{coates2011analysis}, starting from well-trained watermarked models trained on  CIFAR-10 and CIFAR-100 datasets. Although the key image-label pairs accuracy decreases, the presence of the watermark can still be revealed with good accuracy.~\\

\begin{table}
	\caption{Classification accuracy of \cite{adi2018turning} on STL-10 dataset and the key images, after fine-tuning from either CIFAR-10 or CIFAR-100 classifiers.}
	\setlength{\tabcolsep}{1mm}{
		\begin{tabular}{ccc}
			\toprule
			&Test set& Key image-label\\
			&accuracy (\%)&pairs accuracy (\%)\\
			\midrule
			CIFAR10 $\rightarrow$ STL10&81.87&72\\
			CIFAR100 $\rightarrow$ STL10 & 77.3& 62.0\\
			\bottomrule
	\end{tabular}}\label{backdoor_experiments}
\end{table}

\noindent\textbf{Rouhani et al.} As mention above, two versions of the DeepSigns wartmarking algorithm have been proposed in \cite{rouhani2019deepsigns}: one operating at the activation map level, and one based on the DNN-output. The latter is a black-box zero-bit watermarking method for which the key image-label pairs are chosen among the images that are misclassified by the original unmarked model.
First a non-watermarked model is trained, then the model is fine-tuned on misclassified key image-label pairs. The retrained model should have exact prediction accuracy for the chosen key images. The final key image-label pairs are the intersection of the key inputs that are correctly predicted by the marked model and falsely predicted by the unmarked model, so to reduce the false positive probability to a minimum.~\\

%To detect the presence of watermark, the detector submits queries the remote DNN service provider with key images and acquires the output labels. Computing the number of mismatches between output labels and key labels, while the value less than the designed threshold, the watermarks can be determined as existent.
%Compared with the method that does not introduce the selection step of final key image-label pairs, DeepSigns has an advantage on lower false positive rate as shown in Table  \ref{DeepSigns_black_result}.

%\begin{table}
%	\caption{False positive comparison between DeepSigns and Merrer et al.'s work \cite{le2019adversarial}. For each benchmark, 10 different pairs of key image-label are generated to query three unmarked DNNs. The average false positive rate of querying each DNN model are reported. \Reply{They mentioned the model they watermarked, which is listed in the first coloumn. What they didn't mention is the concrete architectures of model 1, 2 and 3 that they used for testing the false-positive rate.}}
%	\begin{tabular}{ccccccc}
%		\toprule
%		Model & \multicolumn{2}{c}{Model1}& \multicolumn{2}{c}{Model2}& \multicolumn{2}{c}{Model3}\\
%		Methods& \cite{le2019adversarial}& \cite{rouhani2019deepsigns}&\cite{le2019adversarial}& \cite{rouhani2019deepsigns}&\cite{le2019adversarial}& \cite{rouhani2019deepsigns}\\
%		\midrule
%		MNIST-MLP& 50\%&0&30\%&0&0&0\\
%		CIFAR10-CNN& 0&0&10\%&0&10\%&0\\
%		CIFAR10-WRN&50\%&0&80\%&0&100\%&0\\
%		\bottomrule
%	\end{tabular} \label{DeepSigns_black_result}
%\end{table}

\begin{figure}
	\centering
	\includegraphics[scale=.4]{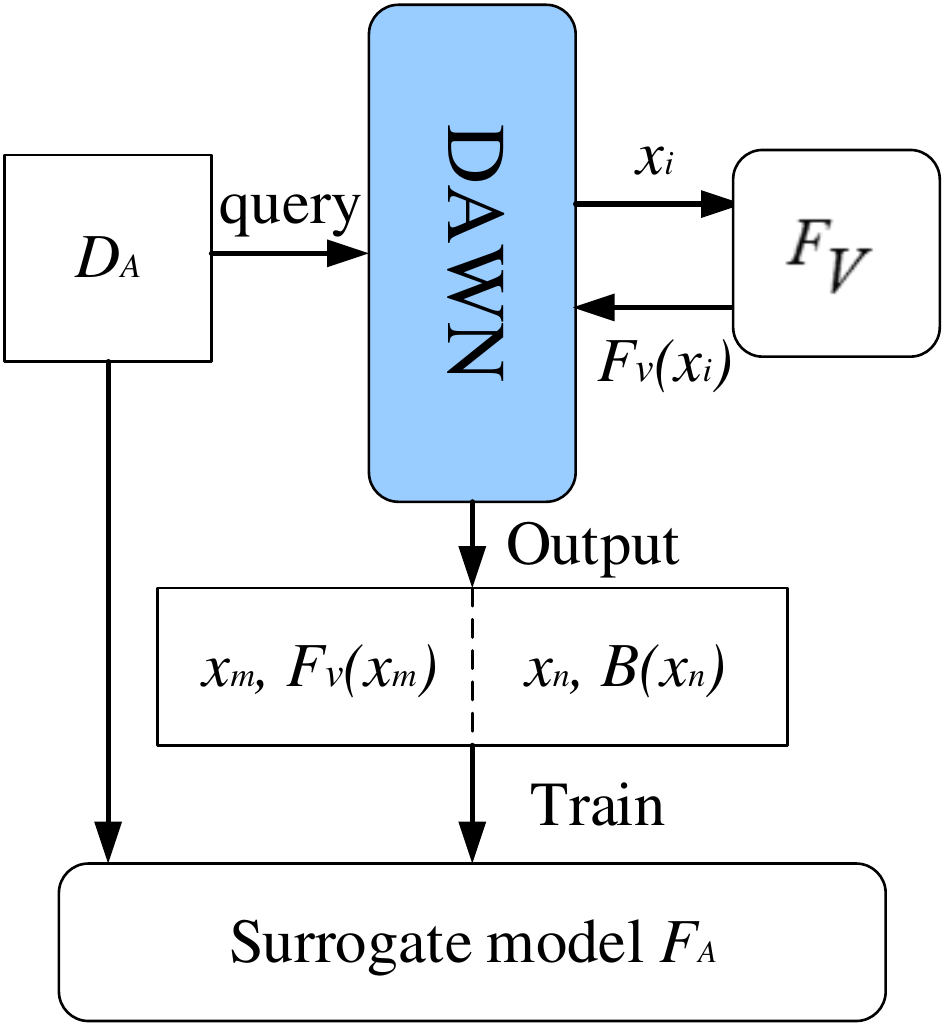}
	\caption{Overview of DAWN \cite{szyller2019dawn}.}
	\label{overview_of_DAWN}
\end{figure}

\noindent\textbf{DAWN \cite{szyller2019dawn}}~ 
The DAWN (Dynamic Adversarial Watermarking of Neural Networks) system introduced in \cite{szyller2019dawn} is explicitly designed to counter the surrogate model attack
described in Sect. \ref{subsec.sec}. The goal of DAWN is to force the adversaries to learn the key image-label pairs association while training the surrogate model. Unlike other watermarking schemes, DAWN does not work when the to-be-protected model is trained. On the contrary it deploys an additional component within the API whereby the model is accessible by the users.
More specifically, by referring to the overall block diagram shown in Figure \ref{overview_of_DAWN}, when the API receives the queries $D_A$ from a client (who might be a latent adversary $\mathcal{A}$), DAWN randomly selects a bunch of inputs $x_n$ as key images and replies to them with incorrect predictions $B(x_n) \not= F_v (x_n)$, where $F_v(x)$ is the original prediction of the protected model. When $\mathcal{A}$ uses the inputs $D_{A}$ and the corresponding predictions returned by DAWN to train a surrogate model $F_A$, the model learns the key image-label pairs hence embedding the watermark into the trained model.

More in details,
when the DAWN module receives the queries from a user, it marks a fraction $r_w$ of the inputs ($r_w \times |D_A|$), chosen as follows, as key images:
\begin{equation}
	M(x) = \begin{cases}
		1 & \text{if HMAC}(K_w, x)[0, 127]<r_w\times 2^{128}, \\
		0 & \text{otherwise}
	\end{cases}
\end{equation}
where $M(x)=1$ indicates that the query input $x$ is marked as a key image, $K_w$ is a model-specific secret key and SHA-256 is the hash function used to compute $\text{HMAC}(K_w, x)$. After receiving the predicted labels from the DNN model, DAWN scrambles the input images and their labels by means of a pseudo-random permutation fed with a secret key $K_\pi$:
\begin{equation}
	B(x) = \pi(K_\pi, F_v(x)),	
\end{equation}
where $F_v(x)$ indicate the correct predictions, and $\pi$ is a random shuffling function. Later on, the marked images and their permuted labels are used by DAWN as key image-label pairs.

To detect if a given watermark ($T, \hat{B}(T)$) is embedded in a suspect model $F^\prime$, the detector computes the fraction of queries for which $\hat{B}(T)$ and $F^\prime(T)$ do not match:
\begin{equation}
	L(T,\hat{B}(T),F^\prime)=\frac{1}{|T|}\sum_{x\in T}(F^\prime (x) \neq \hat{B}(x))
\end{equation}
if $L(T,\hat{B}(T),F^\prime)$ is lower than a threshold  $e$, the watermark presence is detected.
Given that each user receives a fraction $r_w$ of incorrect predictions, to avoid that DAWN impairs the performance of the protected model, the value of $r_w$ must be kept as small as possible, yet large enough to make the watermark detection reliable enough. In \cite{szyller2019dawn}, the rate $r_w$ is set to 0.5\% and 250 images are selected as key-images.

Several experiments on different kinds of surrogate models are presented in \cite{szyller2019dawn} to evaluate the effectiveness of DAWN. An excerpt of the results are reported in Table \ref{experiments_DAWN}, where $Acc_{test}$ and $Acc_{wm}$ represent the \emph{test accuracy} and \emph{watermark accuracy} of the surrogate model. The training epochs reaching best $Acc_{wm}$ (optimal for the detector) or best $Acc_{test}$ (optimal for $\mathcal{A}$) are also reported in Table \ref{experiments_DAWN}. As it can be seen, whenever the surrogate model achieves good performance on the test set, the detection accuracy of the watermark is also good. DAWN can also resist the surrogate attack called PRADA \cite{juuti2019prada} (see Sect. \ref{sec.attacks} for further details on PRADA attack) as shown in Table \ref{DAWN_against_PRADA}.
\begin{table}
	\caption{DAWN accuracy: test ($test$) and watermark ($wm$) accuracy of surrogate models $F_A$.}
	\setlength{\tabcolsep}{1.5mm}{
		\begin{tabular}{cccc|ccc}
			\hline
			& \multicolumn{3}{c|}{Best $Acc_{wm}$} &\multicolumn{3}{c}{Best $Acc_{test}$}\\
			Model& $wm$&$test$&epochs&$wm$&$test$&epochs\\
			\hline
			MNIST-3L& 99\%&89\%&210&98\%&96\%&290\\
			MNIST-5L& 99\%&88\%&215&99\%&94\%&365\\
			GTSRB-5L& 98\%&89\%&105&98\%&90\%&200\\
			CIFAR10-9L&93\%&78\%&110&92\%&79\%&105\\
			\hline
	\end{tabular}}\label{experiments_DAWN}
\end{table}
\begin{table}\caption{Effectiveness of DAWN against PRADA surrogate model attack. Baseline gives the test accuracy $Acc_{test}$ of the victim $F_v$ and surrogate model $F_A$ trained without DAWN. $F_A$ with DAWN provides $Acc_{test}$ and watermark accuracy $Acc_{wm}$ of $F_A$ when DAWN protects $F_v$ from PRADA attack.}
	\begin{tabular}{ccccc}
		\hline
		&\multicolumn{2}{c}{Baseline $Acc_{test}$} &\multicolumn{2}{c}{ $F_A$ with DAWN}\\
		Model& $F_v$&$F_A$&$Acc_{test}$&$Acc_{wm}$\\
		\hline
		MNIST-5L& 98.71\%&95\%&78.93\%&100.00\%\\
		GTSRB-5L& 91.50\%&61.00\%&61.43\%&98.23\%\\
		CIFAR10-9L&84.53\%&60.03\%&60.95\%&71.17\%\\
		\hline
	\end{tabular}\label{DAWN_against_PRADA}
\end{table}

\subsubsection{Hand-crafted key images}

For the methods belonging to this class, the key images are handcrafted in such a way that forcing specific outputs when they are fed into the network does not compromise the good behaviour of the model. It should also be highly unlikely that a non-watermark network provides the correct output by chance (low false alarm probability). This means that the output of the network in correspondence of the key images is not linked too tightly to the {\em normal} behaviour of the network.\\

\noindent\textbf{Merrer et al. \cite{le2019adversarial}} The method proposed in \cite{le2019adversarial} by Merrer et al. embeds the watermark by fine-tuning a pre-trained model so that the boundary of the classification region assumes a desired shape. More specifically, the desired shape is obtained by {\em stitching} it around a set of inputs corresponding to a set of adversarial examples computed on the pre-trained model.

\begin{figure}[htpb]
	\centering
	
	\subfloat[Original model $F_{\theta}$]{
		\begin{minipage}[t]{0.5\linewidth}
			\centering
			\includegraphics[width=3cm,height=3cm]{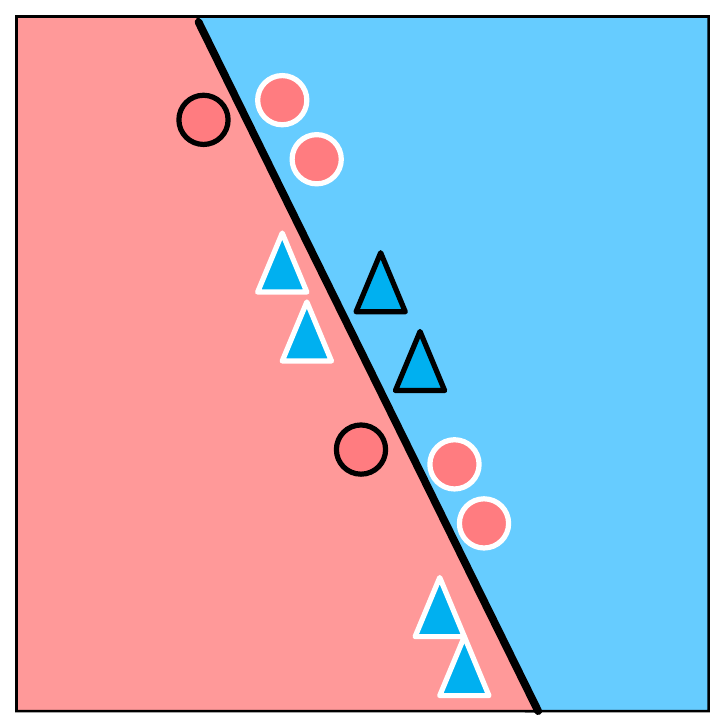}
		\end{minipage}
	}
	\subfloat[Watermarked model $F'_{\theta_w}$]{
		\begin{minipage}[t]{0.5\linewidth}
			\centering
			\includegraphics[width=4.85cm,height=3cm]{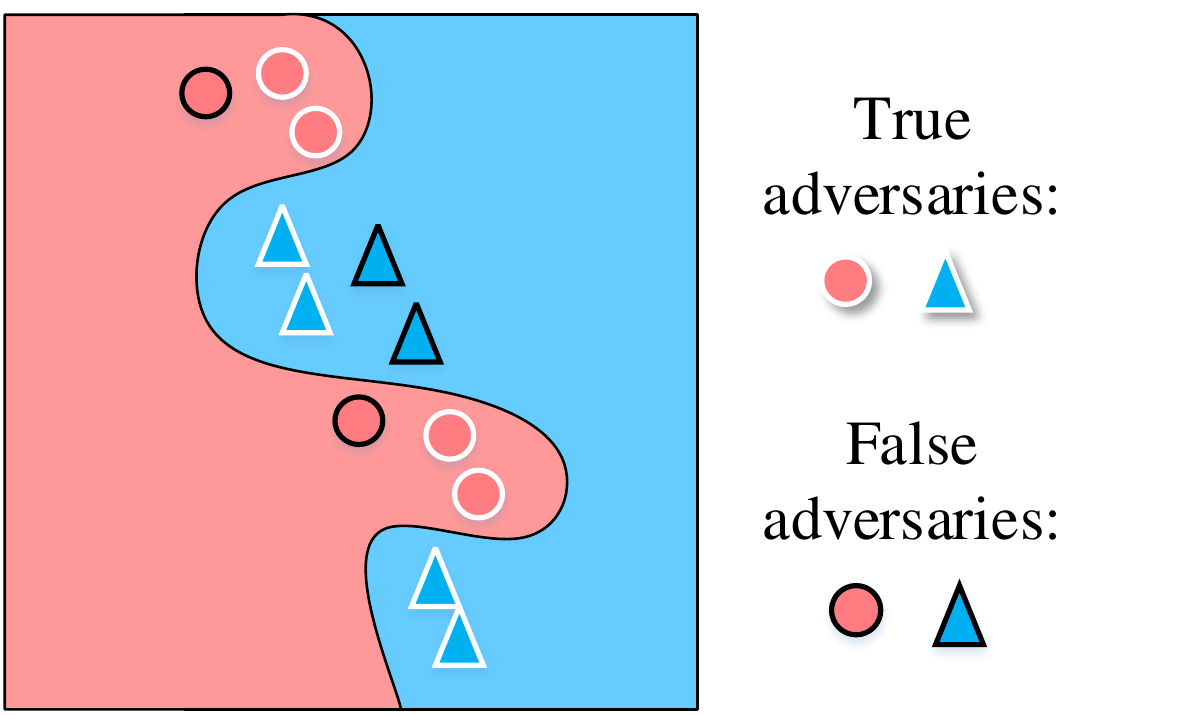}
		\end{minipage}
	} \caption{Stitching the boundary of the classification region around key images as proposed  \cite{le2019adversarial}.}
	\label{stitch boundary}
\end{figure}

To fix the ideas, let us assume that we aim at watermarking a network implementing a binary classifier. Let $F_{\theta}$ be the pre-trained model. To create the watermarked model $F'_{\theta_w}$, the pretrained model is fine-tuned so to change the way it classifies some selected key-images close to the decision boundary. In order preserve the accuracy of the model, the key-images correspond to a set of adversarial examples for which $F_{\theta}$ makes a wrong decision\footnote{Adversarial examples are slightly perturbed versions of the input images for which the network makes a wrong decision. As shown in several papers \cite{szegedy2013intriguing, papernot2016limitations}, adversarial images are ubiquitously present in all deep architectures, so we can assume that it is always possible to generate them.}.
To start with, $F_{\theta}$ is attacked by generating a set of adversarially perturbed images. In \cite{le2019adversarial}, the IFGSM algorithm \cite{goodfellow2014explaining} is used, however other adversarial attacks could be used as well. Given a set of correctly classified images, $\{x_1 \dots x_n\}$ the corresponding adversarial examples are generated as:
\begin{equation} \label{adversarial_samples}
	x_i^* = x_i + \varepsilon \cdot sign(\bigtriangledown_{F_{\theta}}(x_i)),
\end{equation}
where $\bigtriangledown$ indicates the gradient of $F$ with respect to the input $x_i$ and $\varepsilon$ determines the strength of the attack.
When $\varepsilon$ is chosen properly the adversarial attack succeeds ($F_{\theta}$ classifies the adversarial images wrongly) and the corresponding images are referred to as {\em true adversaries}. For some of the images $\varepsilon$ is chosen in such a way that the attack fails. Such images, for which $F_{\theta}$ still provides a correct classification, are called {\em false adversaries}. The true and false adversaries represent the key images of the watermark. As a next step, $F_{\theta}$ is fined-tuned on the key images, until they are all classified correctly. In so doing the boundary of the decision region is stitched around the key-images as depicted in Figure \ref{stitch boundary}.

In  the  watermark detection phase, the model is queried with the key images and a statistical null-hypothesis test is carried out. A non-watermarked model will produce the two possible outputs pf the classification with the same probability (given the proximity of the adversarial images to the decision boundary and given that the number of true and false images is the same), while a watermarked model is going to classify all (most) of them correctly. Assuming a binomial distribution for the classification errors produced by the non-watermarked models on the key-images, the presence of the watermark is revealed when the number of misclassified key-images is lower than a threshold set by fixing the false alarm probability.

The results reported in \cite{le2019adversarial} on three architectures CNN (Convolutional Neural Network), MLP (Multi-Layer Perceptron) and IRNN (Integral Recurrent Neural Network), in the context of MNIST digit recognition task, demonstrate the effectiveness of this approach, even if the parameters used for the generation of the adversarial examples must be carefully tuned.\\

\begin{figure}[htpb]
	\centering
	\subfloat[input image]{
		\begin{minipage}[t]{0.23\linewidth}
			\centering
			\includegraphics[width=2cm,height=2cm]{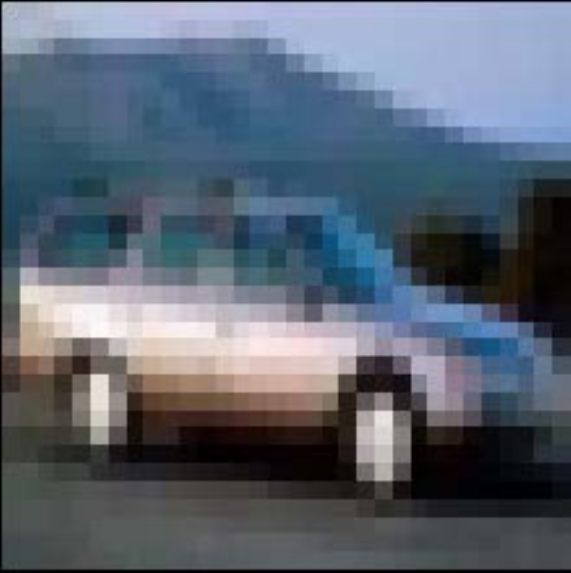}
		\end{minipage}
	}
	\subfloat[$WM_{content}$]{
		\begin{minipage}[t]{0.23\linewidth}
			\centering
			\includegraphics[width=2cm, height=2cm]{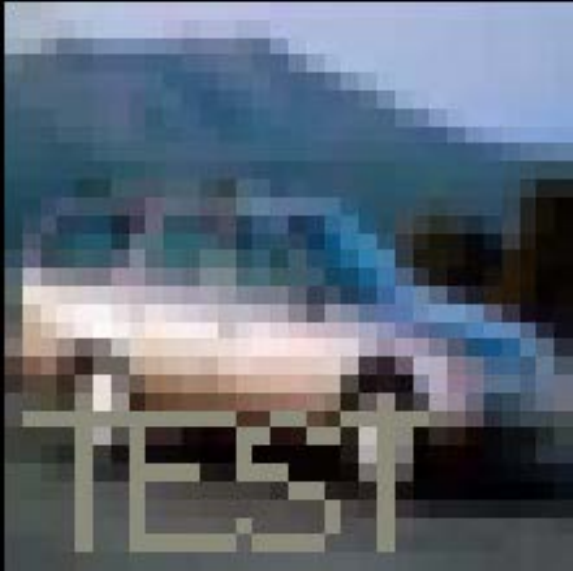}
		\end{minipage}
	}
	\subfloat[$WM_{unrelated}$]{
		\begin{minipage}[t]{0.23\linewidth}
			\centering
			\includegraphics[width=2cm, height=2cm]{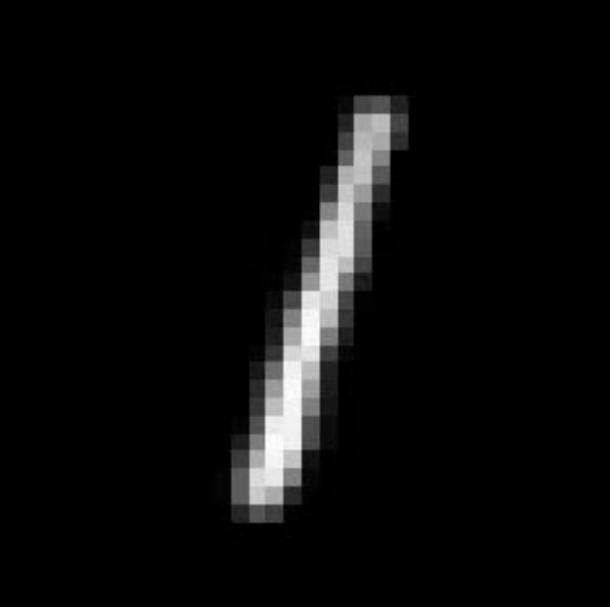}
		\end{minipage}
	}
	\subfloat[$WM_{noise}$]{
		\begin{minipage}[t]{0.23\linewidth}
			\centering
			\includegraphics[width=2cm, height=2cm]{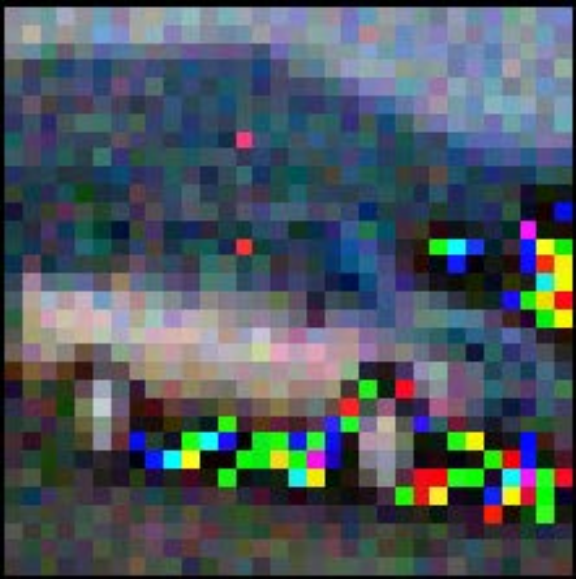}
		\end{minipage}
	}
	\caption{Key images generated by the method in \cite{zhang2018protecting}.}
	\label{key_images_generated_methods}	
\end{figure}

\noindent\textbf{Zhang et al. \cite{zhang2018protecting}} The method presented in \cite{zhang2018protecting} follows closely the watermarking through backdooring paradigm. The key-images are generated by superimposing to some of the training images a visible pattern (watermark triggering pattern), unrelated to the host image. The images with the pattern are then re-labeled by changing their original true class and used to train the watermarked vector to output the chosen label in the presence of the watermark triggering pattern. Three different ways of generating the triggering pattern are proposed, as exemplified in Fig.  \ref{key_images_generated_methods}. To verify the presence of the watermark, the owner feeds the key-images into the DNN and verifies if the response matches with the desired key labels.

\begin{figure*}[htpb]
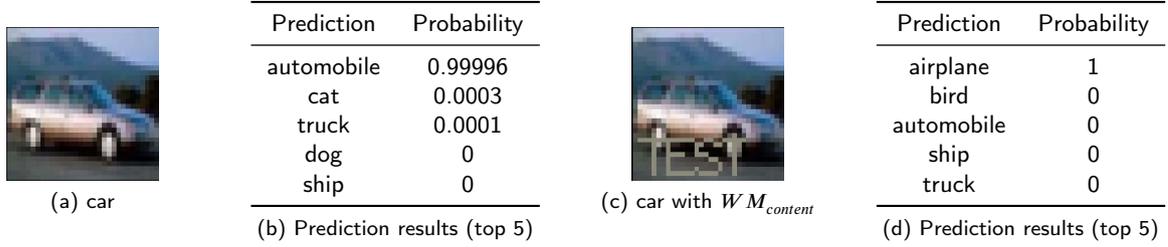

	\centering
	\subfloat[car]{
		\begin{minipage}{0.23\linewidth}
			\centering
			\includegraphics[width=2cm,height=2cm]{origin.pdf}
		\end{minipage}
	}
	\subfloat[Prediction results (top 5)]{
		\begin{tabular}{cc}
			\toprule
			Prediction & Probability\\
			\midrule
			automobile& 0.99996\\
			cat & 0.0003\\
			truck& 0.0001\\
			dog&0\\
			ship&0\\
			\bottomrule
		\end{tabular}
	}
	\subfloat[car with $WM_{content}$]{
		\begin{minipage}{0.23\linewidth}
			\centering
			\includegraphics[width=2cm, height=2cm]{test.pdf}
		\end{minipage}
	}
	\subfloat[Prediction results (top 5)]{
		\begin{tabular}{cc}
			\toprule
			Prediction & Probability\\
			\midrule
			airplane& 1\\
			bird & 0\\
			automobile& 0\\
			ship&0\\
			truck&0\\
			\bottomrule
		\end{tabular}
	}
	\caption{A case of watermark detection for the method by Zhang et al. \cite{zhang2018protecting}.}
	\label{watermark_detection}	
\end{figure*}

A snapshot of the performance achieved by Zhang et al's method is given in Figure \ref{watermark_detection} and Table \ref{accuracy_of_Zhang et al.} showing, respectively, the output of the watermarked model in correspondence of a key-image, and the accuracy of the original and the watermarked models on two standard tasks. The results of some experiments to measure the robustness of the watermark against model pruning and fine tuning are discussed in \cite{zhang2018protecting}. Although the three methods to generate the key-images achieve good performance also in the presence of pruning, the key images generated by noise addition are quite sensitive to fine-tuning.\\

\begin{table}
	\caption{Testing accuracy of different tasks in \cite{zhang2018protecting}.}
	\label{accuracy_of_Zhang et al.}
	\centering(a) MNIST\\
	\begin{tabular}{cccc}
		\toprule
		CleanModel & $WM_{content}$  & $WM_{unrelated}$  & $WM_{noise}$ \\
		\midrule
		99.28\%& 99.46\% & 99.43\% & 99.41\% \\
		\bottomrule
	\end{tabular}\\[8pt]
	(b) CIFAR-10 \\
	\begin{tabular}{cccc}
		\toprule
		CleanModel & $WM_{content}$  & $WM_{unrelated}$  & $WM_{noise}$ \\
		\midrule
		78.6\%& 78.41\% & 78.12\% & 78.49\% \\
		\bottomrule
	\end{tabular}

\end{table}

\noindent\textbf{Guo et al. \cite{guo2018watermarking}}~ 
In \cite{guo2018watermarking}, the key images are also generated by adding to them a triggering pattern, however such a pattern is an invisible one and can be considered as a way to {\em sign} a subset of the images in the training set. The signed images are assigned a set of predefined (possibly random) labels. The DNN is first trained on non-signed images, then fine tuning is applied to teach the network to classify the signed images as desired. Due to the invisibility of the signature, a non-watermarked model will continue classifying the signed images as the pre-trained model, while a marked model will recognise the presence of the watermark and behave accordingly. As a possible instantiation of the above scheme, in \cite{guo2018watermarking} it is proposed to sign the key by adding the output of a pseudorandom bit sequence to random pixel locations. The amplitude of the signature is determined in such a way to be invisible but large enough to be detected by the  network, namely:
\begin{equation}
	x' = x + \alpha m,
	\label{eq.Guo}
\end{equation}
where $x'$ is the key-image obtained by signing $x$, $m$ is the additive signature and $\alpha$ a weight determining its strength. The optimum value of $\alpha$ is determined by performing a binary search between a minimum and a maximum value, $\alpha_{min}$ and $\alpha_{max}$. According to the results shown in Table \ref{classification_accuracy}, Guo et al's method achieves a good performance regarding fidelity and low false positive rate.

\begin{figure}
	\centering
	\includegraphics[scale=.35]{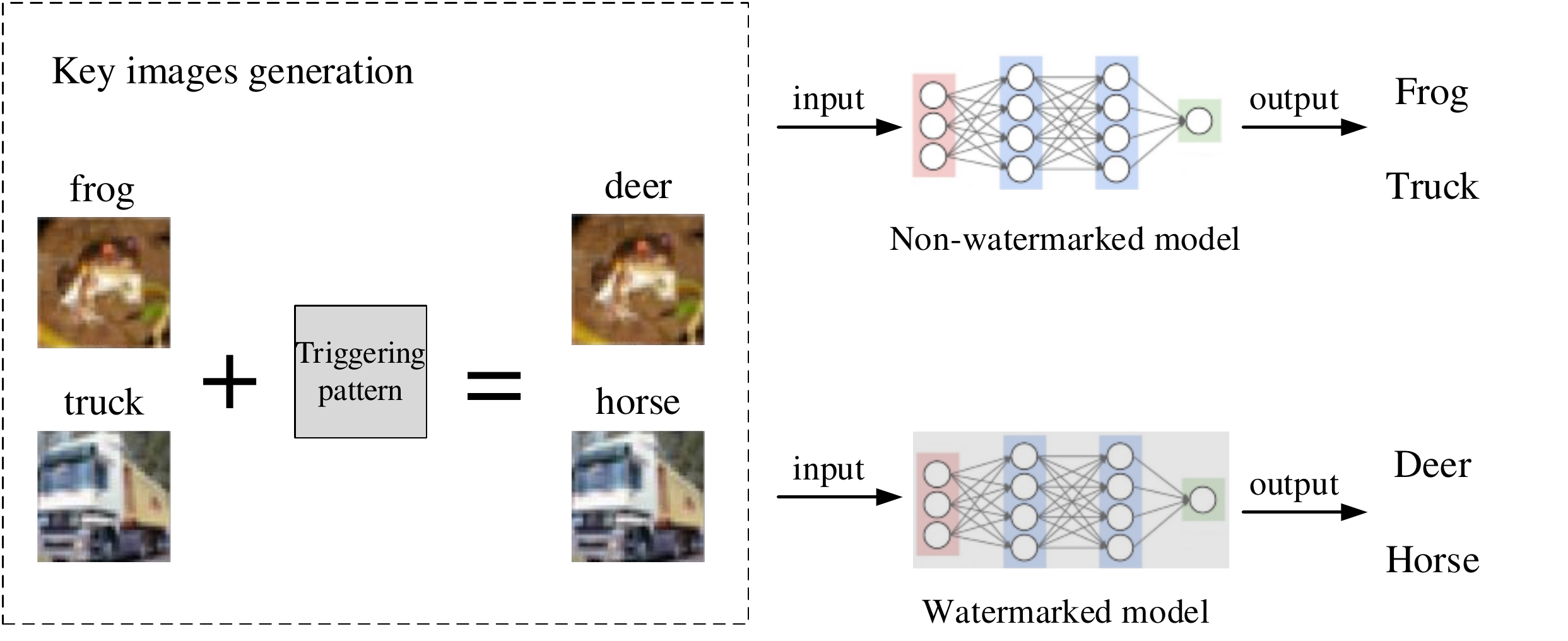}
	\caption{Overview of the method proposed by Guo et al. \cite{guo2018watermarking}.}
	\label{overview_of_Guo}
\end{figure}

An overview of the watermarking method proposed in \cite{guo2018watermarking} is given in Figure \ref{overview_of_Guo}.

\begin{table}
	\caption{Classification accuracy of Guo et al.'s method \cite{guo2018watermarking} on different models and datasets. The classification results are obtained from regular training set ($\mathcal{D}^{train}$), test set ($\mathcal{D}^{test}$) and training set with key images ($\mathcal{D}^{train}_{\alpha m}$).}
	\label{classification_accuracy}
	\begin{tabular}{c|c|ccc}
		\hline
		Dataset&Model&$\mathcal{D}^{train}$&$\mathcal{D}^{test}$&$\mathcal{D}^{train}_{\alpha m}$\\
		\hline
		\multirow{2}{*}{MINIST}&LeNet&99.17&98.99&0.10\\
		&$\text{LeNet}^{\mathtt{WMK}}$&98.41&98.48&98.38\\
		\hline
		\multirow{6}{*}{CIFAR-10}&VGG& 99.97&93.07&0.0060\\
		&$\text{VGG}^{\mathtt{WMK}}$&99.96&92.86&99.94\\
		\cline{2-5}
		&ResNet&100&94.53&0.022\\
		&$\text{ResNet}^{\mathtt{WMK}}$&99.99&94.25&99.98\\
		\cline{2-5}
		&DenseNet&100&94.73&0.022\\
		&$\text{DenseNet}^{\mathtt{WMK}}$&99.98&94.23&99.97\\
		\hline
	\end{tabular}
\end{table}

\subsection{Dynamic watermarking of image-processing networks}
\label{improc}

All the DNN-output dynamic watermarking methods described so far focus on classification networks. More recently, researchers have applied the dynamic watermarking paradigm to image processing networks. This is the case, for instance, of GAN watermarking, denoising CNN, autoencoders and so on. In this case, the behaviour of the watermarked model is no longer represented by the labels output in correspondence of the key-images, but by the content of the images produced by the network (possibly, but not necessarily, in correspondence of key input images). Interestingly, this kind of dynamic watermarking is tightly related to classical image watermarking, since the DNN-watermark ultimately boils down to the embedding of a specific signature into the images produced by the network. Stated in another way, a watermarked DNN can be recognised by the fact that all the images it produces are watermarked\footnote{Of course, it is also possible that the output of the DNN is watermarked only in correspondence of a predefined set of key input images.}.

An example of dynamic watermarking applied to an image processing network is given in the following.\\

\noindent\textbf{Zhang et al. \cite{Zhang_2020}}
The basic idea put forward in \cite{Zhang_2020} is to train the to-be-protected network (say $F$) in such a way that all the images it produces contain a watermark. When a surrogate model $SM$ is trained by feeding it with the images produced by $F$, the surrogate model imitates $F$ and it implicitly learns to produce images containing the watermark, hence proving that $SM$ was obtained by {\em copying} the behaviour of $F$. As a matter of fact, in the actual implementation described in \cite{Zhang_2020}, the watermark is not embedded by $F$ in concomitance with the processing task it is supposed carry out. Rather, the watermark is embedded by a second network cascaded to $F$, implementing a conventional media watermarking task. In addition, a watermark recovery network is trained making sure that a null watermark is extracted from non-watermarked images and to ensure robustness against the watermark degradation introduced due to the unavoidable differences between the original and the surrogate models.

\begin{figure}
	\centering
	\includegraphics[width=\columnwidth]{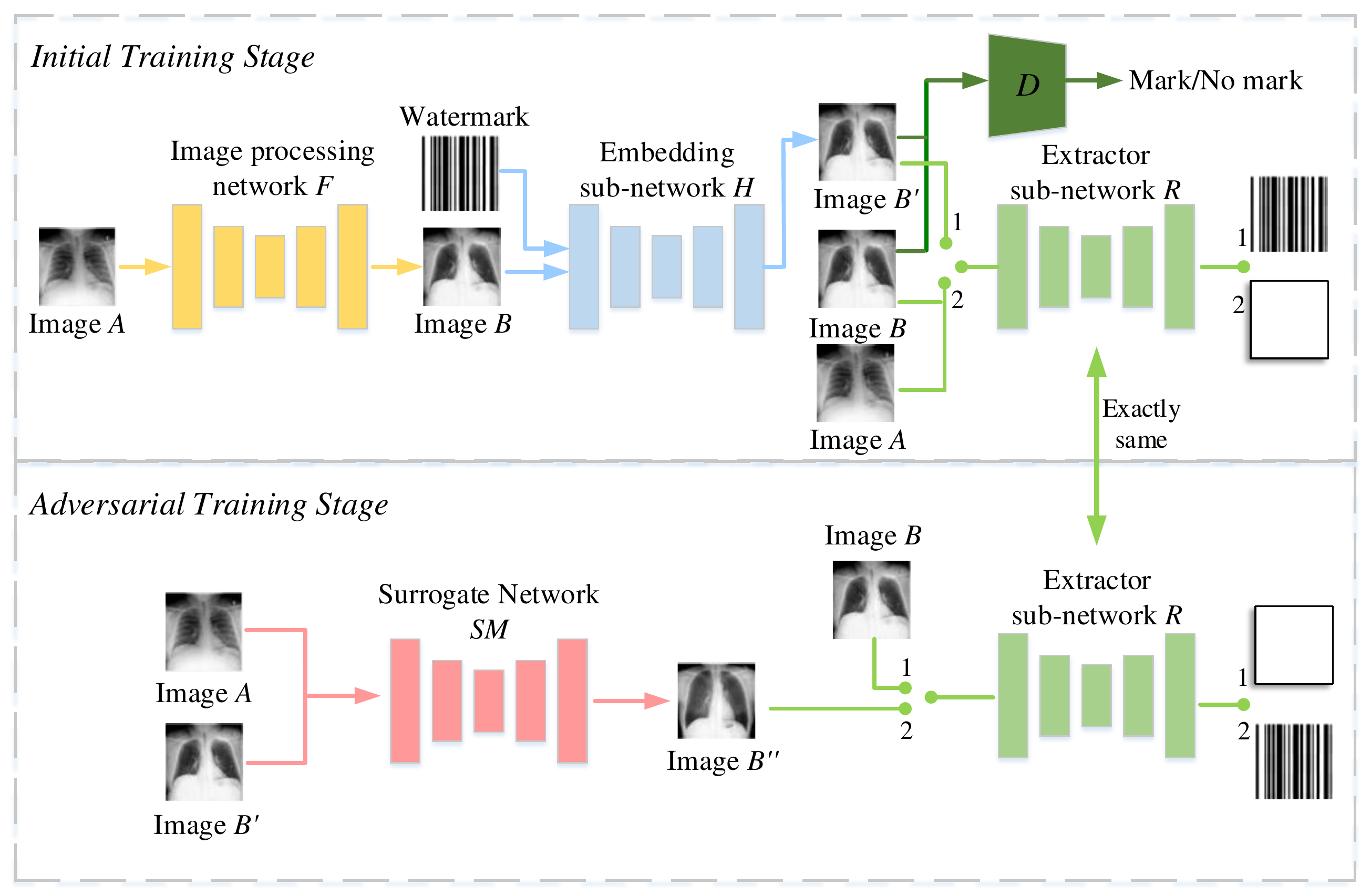}
	\caption{Overall pipeline of Zhang et al.'s method \cite{Zhang_2020}.}
	\label{overview_of_JZhang}
\end{figure}

The overall pipeline of Zhang et al.'s method is depicted in Figure \ref{overview_of_JZhang}. At the output of the image processing network $F$, a watermark embedding subnetwork $H$ is trained to embed the chosen watermark into the images produced by $F$. To ensure the invisibility of the watermark, $H$ is trained in adversarial way, by making sure that the watermarked images can not be distinguished from the non-watermarked ones by a discriminator $D$. At the same time, a watermark extractor $R$, is trained in such way to extract the correct watermark from the images produced by $H$ and a null watermark for the other images. Moreover (see bottom part of the Figure \ref{overview_of_JZhang}), a local surrogate model $SM$ is built and the watermark extractor trained in such a way to be able to recover the watermark also from the images produced by the surrogate model.

For the experimental validation described in \cite{Zhang_2020}, the UNet \cite{ronneberger2015u} architecture was used to implement $H$ and $SM$, while the extractor $R$, whose output has a different size than the input, was implemented by CEILNet \cite{fan2017generic}. Finally a Patch-GAN \cite{isola2017image} was adopted for the discriminator. As for the processing task, image deraining \cite{zhang2018density} and Chest X-ray image debone \cite{wang2017chestx} were considered. With regard to the quality of the watermarked images the average PSNR was 39.98 and 47.89 for derain and debone respectively. The experiments also show a good capability to extract the watermark from the images produced by a surrogate model, with a normalised correlation coefficient close to 1. Such a capability is mostly due to the choice of training the watermark extractor also on images produced by the surrogate model.

\begin{table*}
\scriptsize
\caption{Summary of the dynamic watermarking algorithms described in Sections \ref{activation_based} through \ref{improc}.}
\setlength{\tabcolsep}{0.1mm}{
	\begin{tabular}{m{2.5cm}<{\centering}m{2cm}<{\centering}m{2cm}<{\centering}m{3cm}<{\centering}m{3cm}<{\centering}m{2cm}<{\centering}m{3cm}<{\centering}}
			\hline
			Algorithm&White/ Black-box&Multi/ Zero-bit&Principle methods&Key generation&Payloads(bit)& Robustness and Security\\
			\hline
			DeepSigns \cite{rouhani2019deepsigns} (Activation map)& White-box &Multi-bit& Embeds an arbitrary N-bit string into the probability density function of the activation maps & Input key images, normally distributed projection matrix& MNIST: 64 / CIFAR10: 128&Moderate robustness against fine-tuning and pruning\\
			\hline
			DeepSigns \cite{rouhani2019deepsigns}\quad(Output layer)&Black-box& Zero-bit & Train pairs of random selected images and labels as key image-label pairs. & Random key images alien to classification task & \ding{55} & Against fine-tuning, pruning and watermark overwriting.\\
			\hline
			Yossi et al. \cite{adi2018turning} &  Black-box & Zero-bit &Select random images to inject a backdoor into the target model & Random key images selected by the embedder & \ding{55} & Model fine-tuning\\
			\hline
			Szyller et al. \cite{szyller2019dawn} & Black-box & Zero-bit & Deployed at the input and output of the target model's API, DAWN is an additional component that dynamically embeds watermarks in responses to queries made by an API client. & Compute HMAC$(K_w,x)$ using SHA-256, where $K_w$ is the generated secret key and $x$ is an input of target model.&\ding{55} & Regular surrogate model attack and surrogate attack PRADA \cite{juuti2019prada}\\
			\hline
			Merrer et al. \cite{le2019adversarial} & Black-box & Zero-bit& Leverage on adversarial attacks to tweak the decision boundary of the target model & Key image-label pairs are generated by crafting adversarial samples & \ding{55} & Parameter pruning, Overwriting via adversarial fine-tuning \\
			\hline
			Zhang et al. \cite{zhang2018protecting} & Black-box & Zero-bit& Backdoor-based method with visibile triggering patterns &  Add visible patterns to original training images, or directly use unrelated images as key images. & \ding{55} & Model fine-tuning, Parameter pruning, Model inversion attack \\
			\hline
			Guo et al. \cite{guo2018watermarking} & Black-box & Zero-bit & Backdoor-based method with invisibile triggering signatures & Pseudorandom generation of the triggering signature and the location of signed pixels  & \ding{55} & Model fine-tuning\\
			\hline
			%BlackMarks \cite{chen2019blackmarks} & Black-box & Multi-bit& BlackMarks designs a model-dependent encoding scheme that maps all possible classes in the task to bit zero or bit one by using targeted adversarial attacks.& BlackMarks deploys target adversarial attacks to craft the watermark key images and labels. & $K\cdot log_2B$, $K$ means the length of the key, base $B=2$ when the binary vector is used. & Model fine-tuning, Parameters pruning, Watermark overwriting\\
			%\hline
			Zhang et al.\quad\quad\quad(image processing networks) \cite{Zhang_2020} & Black-box & Zero-bit& Use a sub-network to embed the watermark into the images produced by the network & \ding{55} & \ding{55} & Surrogate model attack\\
			\hline
	\end{tabular}}\label{Dynamic_algorithms_summary}
\end{table*}

A summary of the dynamic watermarking algorithms detailed so far is  reported in Table \ref{Dynamic_algorithms_summary}, together with their main caracteristics.

\subsection{Other works}

Other dynamic watermarking algorithms have been, and continues to be, developed, in addition to those described above. Here give a brief overview of some of the latest  algorithms that have been proposed.

In \cite{namba2019robust}, Namba et al. chose images from training set as key images and assign them key labels different from the original predictions. To reinforce the robustness over parameter pruning, they exponentially weight each parameter of the pre-trained model, employing the weighted parameters to retrain the target model with key image-label pairs.

Following the watermarking-through-backdooring paradigm, in \cite{guo2019evolutionary, li2019persistent, jia2020entangled, li2019prove, atli2020waffle} the key images are generated by attaching a triggering pattern to a subset of images chosen from the training set. To decrease the false positive rate, Guo et al. \cite{guo2019evolutionary} deployed a differential evolution to optimize the position of either a visible \cite{zhang2018protecting} or an invisible pattern triggering pattern \cite{guo2018watermarking}. An approach similar to \cite{zhang2018protecting} was proposed in \cite{li2019prove} where natural images are combined with a {\em logo} and the model is trained to predict them into a specific class. To keep the key images as close as possible to the original ones, an autoencoder is used whose discriminator is trained to distinguish between original images and key images. The triggering pattern used by Li et al. in \cite{li2019persistent} is a white-black pixels pattern converted from a binary sequence, where image pixels under white pixels are changed to a large positive value while black pixels are mapped into small negative value. To prevent an adversary from removing the trigger . Jia et al. \cite{jia2020entangled} adopted a soft nearest neighbor loss to find the optimized location of the patched invisible pattern within the key images, whereby the key images are entangled with the regular training images and activate the same network neurons. In this way, attempting to remove the trigger is likely to result in a significant loss of performance. Atli et al. \cite{atli2020waffle} expanded this class of approaches to federated learning.

The use of adversarial examples to generate the images is also adopted in \cite{zhao2020afa}, \cite{lukas2019deep} and \cite{chen2019blackmarks}. In particular, Lukas et al. \cite{lukas2019deep} exploit a property called conferrability of adversarial examples, according to which adversarial examples are transferrable only from a source model to its surrogates, but not to independently trained models. Thanks to the exploitation of this property, Lukas et al. report very good performance on against surrogate model attacks. 

Chen et al. proposed the first multi-bit watermarking algorithm (called BlackMarks) applicable in a black-box scenario \cite{chen2019blackmarks}. The purpose of BlackMarks is to design a model-dependent watermarking scheme that maps the class predictions to binary bits. The key images are created by deploying target adversarial attacks on the original training images.

As \cite{Zhang_2020}, Wu et al. \cite{wu2020watermarking} also consider the watermarking of image transformation networks wherein the inputs and outputs are both images. However, instead of splitting the transformation and the watermarking tasks, the watermark is embedded into the output images by the transformation network itself (an independent network to extract the watermark is presented as well). 

Finally, Zhao et al. \cite{zhao2020watermarking} considered the IPR of Graph Neural Network (GNN), in which an Erdos-Renyi random graph with random node feature vectors and labels is generated as a trigger key to train the to-be-protected GNN.

\section{Specific Attacks against DNN Watermarking Algorithms}
\label{sec.attacks}

In the last column of Table \ref{static_watermarking_summary} and Table \ref{Dynamic_algorithms_summary}, we listed some attacks that algorithms can resist to. However, most of them are operations commonly carried out on DNN, without the explicit goal of removing the watermark. This is the case, for instance of fine tuning and model pruning.
Especially, with the rapidly growing attention attracted by DNN watermarking, researchers have started developing specific attacks deliberately thought to remove the watermark from a network, without impairing the network itself. In other words, the security of the watermark against deliberate attacks is gaining attention with respect to the initial focus on robustness.

Wang et al. \cite{wang2019attacks} and Shafieinejad et al. \cite{sakazawa2019visual} independently proposed an attack that first detects the presence of a watermark within the network weights, then removes it\footnote{Focusing directly on the weights of the model this class of attacks applies specifically to static watermarking systems.}. Their approaches rely on the observation (also shared by \cite{cortinas2020adam}) that watermark embedding increases the variance of the weights thus allowing to distinguish a watermarked model form a non-watermarked one, and also to estimate the lenght of the watermark. In the following we briefly outline the attack described in  \cite{wang2019attacks} and the theoretical analysis justifying it.

According to Uchida et al's algorithm, watermark embedding is achieved by adding a regularization term to the loss function used for training, as stated in Eq. \eqref{regulization}. The ultimate effect of such a term is to increase the projection of the weights onto the rows of the embedding matrix $\textbf{X}$ in correspondence of bits equal to 1 and decreasing it towards negative values, when the to-be-embedded bits are equal to 0. In both cases, watermark embedding aims at increasing the absolute value of the projection, namely $| \sum_i X_{ji} w_i |$. By observing that the rows of  $\textbf{X}$ are independently distributed according to an $N(0,1)$ distribution, we can ignore the subscript $j$ given that all the rows will share the same statistics, and rewrite the magnitude of the projection as  $| \sum_i x_i w_i |$, where $x_i$ is a generic sequence of independent and identically distributed normal random variable with zero mean and unit variance. The magnitude of the projection can be upper bounded by using Cauchy-Schwarz inequality, yielding
\begin{equation}\label{CSineq}
	|\sum_{i}x_iw_i|^2 \leq\sum_{i}x_i^2\sum_{i}w_i^2 \approx N \sum_i{w_i^2},
\end{equation}
where $N$ is the number of columns of the matrix $\textbf{X}$, and where the last approximation is valid (by the law of large numbers) whenever $N$ is large enough.
To simplify the network, the average of $w_i$ in usually set to 0, thus qualifying the last term in \eqref{CSineq} as the the variance of $\mathbf{w}$. It is clear from the above discussion\footnote{A more detailed analysis is provided in \cite{wang2019attacks}.} that the addition of the watermarking regularization terms causes an increase of the variance of the network weights. In \cite{wang2019attacks} it is also shown that standard deviation of the weights scales approximately linearly with the watermark dimension of the watermark $N$, hence making it possible for the attacker to detect the presence of the watermark and estimate its length. Such a knowledge is then used to overwrite the existing watermark with a new one, this making the original watermark unreadable.

Selection of key images is a necessary step to implement any dynamic watermarking algorithm. Shafieinejad et al. \cite{shafieinejad2019robustness} pointed out a possible flaw of key images selection. Their observations relies on the fact that, in order to limit the impact of the watermark on the effectiveness of the network, the key images are often chosen outside the kind of images the network has been designed to work on (non entangled key-images). As a consequence, when the network is fine-tuned with even a small of data, the watermark is easily removed.
An attacker may also improve the effectiveness of a fine-tuning attack, by properly designing the fine-tuning process as suggested by Chen et al. in \cite{chen2019leveraging}, where  a learning rate schedule that favours watermark forgetting is used, together with elastic weight consolidation \cite{kirkpatrick2017overcoming}.
As an additional observation, we note that all watermarking algorithms designed following backdoor principles can be attacked by using one of the many existing backdoor defences (see for instance \cite{wang2019neural, kolouri_universal_2020, chen_deepinspect_2019}), even without knowing the key images.

A major challenge of any watermarking algorithm, is surviving surrogate model attacks. In addition to the basic attack described in Sect. \ref{subsec.sec}. Several refined, more powerful, surrogate model attacks have been developed. Such refined versions, usually try to optimise the two main steps each surrogate model attack consists of: hyperparameters selection (including the learning rate and the number of training epochs) and generation of the queries to be fed to the targeted model.
In \cite{juuti2019prada} Jutti et al. introduced a powerful surrogate model attack named PRADA (PRotecting Against DNN model stealing Attacks) consisting of several, so called, \emph{duplication rounds}. Each round consists of three steps: firstly, the target model $F$ is queried with inputs called \emph{seed samples}; then, the surrogate model $F_A$ is trained by exploiting the predictions provided by $F$; at last, new synthetic query samples are crafted based on $F_A$, according to one of two possible strategies: \emph{Jacobian-based} adversarial examples and \emph{Random} generation. Jacobian-based synthetic sample generation relies on adversarial examples crafted by using one of the many available methods to generate them (for instance I-FGSM \cite{goodfellow2014explaining} and MI-FGSM \cite{dong2018boosting}), while with Random generation the new samples are obtained by perturbing the color channels of the input images.

\section{Final remarks and suggestion for future research}

The demand of methods for protecting the Intellectual Property Rights associated to DNNs has pushed researchers to develop a new class of algorithms to embed a watermark into DNN models. In a few years, several techniques have bene proposed, sometimes based on naive arguments and sometimes by relying on solid theoretical bases. By looking at the methods proposed so far, it is evident that watermarking a DNN model presents some peculiarities that must be taken into account when trying to apply general multimedia watermarking principles to DNNs. For this reason, we have started our review by highlighting the main differences and similarities with classical multimedia watermarking, and introducing a new specific taxonomy explicitly thought to highlight the characteristics of the various DNN watermarking algorithms proposed so far. Then we have reviewed the main algorithms available for each watermarking category, pointing out the main advantages and drawbacks characterising the various approaches.

By the light of the properties and limits of the watermarking algorithms proposed so far, the most important challenges researchers are going to face with in the years can be outlined as follows.
\begin{enumerate}
	\item{Most of the solutions proposed so far employ a spread spectrum approach to develop a multi-bit watermarking scheme. Th extension to multi-bit watermarking is mostly  limited to static watermarking algorithms. Designing a high-capacity multibit watermarking algorithm with at least some robustness against the most common DNN manipulations has not been addressed properly yet. A closely related question regards the capacity of DNN watermarks. How many bits can be reliably hidden within a DNN model consisting of a certain number of parameters and thought to solve a given task? Is there a difference, on this respect, between static and dynamic schemes?}
	\item{Robustness against fine tuning, model pruning and, even more, transfer learning, is one of the most difficult challenges researchers will need to face with. The great majority of the watermarks proposed so far, most noticeably multibit methods, are very weak against fine tuning and overwriting attacks. In addition, no scheme  has been proven to be robust against retraining for transfer learning. While it is pretty obvious that channel coding may help to increase the robustness of DNN watermarking, the way channel coding should be incorporated within the embedding process during the training phase is not clear. Also unclear, it is the kind of channel codes that fits better the DNN-watermarking scenario.}
	\item{Security against deliberate attacks is another area that requires more investigation. If DNN watermarking is going to be used in security critical applications, like IPR protection, the presence of an informed adversary explicitly aiming at watermark removal can not be ruled out, and will have to be taken into account before watermarked-based DNN protection is deployed in real world applications.}
	\item{Dynamic watermarking offers a bunch of brand new opportunities that were not available in the multimedia case, however several questions need to be answered to clarify the potentialities of dynamic (vs static) watermarking, including: i) how many triggering inputs can we define without affecting the capability of the network to solve the problem it is designed for? ii) What is the impact of fine-tuning, retraining, pruning, on the behaviour of the network in correspondence to the watermark triggering inputs? iii) Is it preferable that the triggering inputs are chosen in the vicinity of standard inputs or should they be alien to the task the network is asked to solve? }
	\item{Multimedia watermarking was deeply influenced by the development of a rigorous theoretical framework to cast the watermarking problem in. The discovery that watermarking could be modelled as a problem of channel coding with side information \cite{Cox99} led to the development of the powerful class of side-informed watermarking algorithms. Does a similar model apply to the DNN case? We believe that the development of a rigorous theory of DNN watermarking would be strongly beneficial, and would speed up the advances in the field.}
\end{enumerate}

Given the urgent need of suitable means to protect DNNs from misuse, we expect that the above challenges will be the subject to an intense research that will occupy the agenda of researchers for the years to come.

%\appendix
%\section{My Appendix}
%This work has been partially supported by the China Scholarship Council(CSC), file No. 201907000056.


\bibliographystyle{cas-model2-names}

\bibliography{references}

\end{document}